\documentclass[a4paper,11pt]{article}
\pdfoutput=1 

\usepackage{jcappub} 
\usepackage[T1]{fontenc} 
\usepackage{amsbsy}
\usepackage{color}
\usepackage{subcaption}

\usepackage[normalem]{ulem}

\newcommand{\I}{\mathcal I}
\newcommand{\T}{\mathcal T}
\newcommand{\R}{\mathcal R}

\title{\boldmath Relativistic second-order initial conditions for simulations of large-scale structure}

\author[a]{Julian Adamek,}
\author[b]{Juan Calles,}
\author[c,d]{Thomas Montandon,}
\author[b]{Jorge Nore\~na,}
\author[c,e]{Cl\'ement Stahl}

\affiliation[a]{Institute for Computational Science, Universit\"at Z\"urich, Winterthurerstrasse 190, 8057 Z\"urich, Switzerland}
\affiliation[b]{Instituto de F\'{\i}sica, Pontificia Universidad Cat\'{o}lica de Valpara\'{\i}so, Casilla 4950, Valpara\'{\i}so, Chile}
\affiliation[c]{Universit\'e de Paris, CNRS, Astroparticule et Cosmologie, 75006 Paris, France}
\affiliation[d]{Universit\'e Paris-Saclay, CNRS/IN2P3, IJCLab, 91405 Orsay, France}
\affiliation[e]{Universit\'e de Strasbourg, CNRS, Observatoire astronomique de Strasbourg, UMR 7550, 67000 Strasbourg, France}

\emailAdd{julian.adamek@uzh.ch}
\emailAdd{juan.calles.h@mail.pucv.cl}
\emailAdd{thomas.montandon@apc.in2p3.fr}
\emailAdd{jorge.norena@pucv.cl}
\emailAdd{clement.stahl@unistra.fr}

\abstract{Relativistic corrections to the evolution of structure can be used to test general relativity on cosmological scales. They are also a well-known systematic contamination in the search for a primordial non-Gaussian signal. We present a numerical framework to generate RELativistic second-order Initial Conditions (\texttt{RELIC}) based on a generic (not necessarily separable) second-order kernel for the density perturbations. In order to keep the time complexity manageable we introduce a scale cut that separates long and short scales, and neglect the ``short-short'' coupling that will eventually be swamped by uncontrollable higher-order effects. To test our approach, we use the second-order Einstein-Boltzmann code \texttt{SONG} to provide the numerical second-order kernel in a $\Lambda$CDM model, and we demonstrate that the realisations generated by \texttt{RELIC} reproduce the bispectra well whenever at least one of the scales is a ``long'' mode. We then present a generic algorithm that takes a perturbed density field as an input and provides particle initial data that matches this input to arbitrary order in perturbations for a given particle-mesh scheme. We implement this algorithm in the relativistic N-body code \texttt{gevolution} to demonstrate how our framework can be used to set precise initial conditions for cosmological simulations of large-scale structure.}

\begin{document}
\maketitle
\flushbottom

\section{Introduction}\label{sec:Introduction}

We can learn much about the physics of inflation from the statistical properties of cosmological fluctuations. In particular, if these fluctuations are sourced by more than one light degree of freedom, there should be a physical correlation among long and short scales that is absent in all single-field models \cite{Maldacena:2002vr,Creminelli:2004yq,Creminelli:2011rh,Creminelli:2012ed, Pajer:2013ana}. This correlation shows up as a divergence in the so-called squeezed limit of the primordial bispectrum, $B(k_1, k_2, k_3)$, when one of the modes is much smaller than the other two, $k_1 \ll k_2, k_3$. This type of divergence can be constrained by comparing data with the ``local template'' (see e.g.\ \cite{Gangui:1993tt, Komatsu:2010hc}), with amplitude $f_{\mathrm{NL}}$. The study of primordial correlation functions beyond two-point statistics is called primordial non-Gaussianity (PNG). 

The current constraint, given by \textit{Planck}, is $f_{\textrm{NL}} = -0.9 \pm 5.1$ \cite{Akrami:2019izv}. Recent measurements have constrained $f_{\mathrm{NL}}$ using the the scale-dependent halo bias in the power spectrum and obtained $f_{\textrm{NL}} = -12 \pm 21$ \cite{Mueller:2021tqa}.
Future large-scale survey experiments such as \textit{Euclid}, the \textit{Vera Rubin Observatory}, \textit{SKA} or \textit{SPHEREx} \cite{Amendola:2016saw,Zhan:2017uwu,SKA:2018ckk,Dore:2014cca}, are expected to reach percent accuracy for $k \leq 1 \text{ Mpc}^{-1}$ \cite{Schneider:2015yka} which could add more constraints on the primordial scenarios \cite{Biagetti:2019bnp}. In particular, it has been shown in Refs.\ \cite{Desjacques:2016bnm,Dore:2014cca,Karagiannis:2018jdt,Karagiannis:2019jjx} that these surveys could improve the constraints on $f_{\textrm{NL}}$ with a standard deviation $\sigma_{f_{\textrm{NL}}}\sim 1$ for \textit{Euclid} and $\sigma_{f_{\textrm{NL}}}\sim 0.1$ for \textit{SPHEREx} and \textit{SKA}. Most of this constraining power is expected to come from the effect of PNG on galaxy biasing, visible in the large-scale power spectrum, and the bispectrum at large scales or squeezed configurations.

The standard approach to structure formation uses a Newtonian framework, see Ref.\ \cite{Bernardeau:2001qr} for a review. Much work has been done in that framework to describe the non-linear evolution of matter under gravity, especially important at small scales. This is valid when we consider the dynamics of sufficiently small patches of the Universe. Close to the horizon scale relativistic effects become important, as these typically scale as $\mathcal H^2/k^2$ to leading order. Since constraints on PNG come from the large scales, these effects are particularly relevant.  Furthermore, galaxy biasing is a very non-linear phenomenon, and the squeezed bispectrum accounts for the coupling of large and small scales. Thus, one needs a description which is both relativistic and non-linear in order to model them accurately.

Many studies have been performed over the last 30 years to account for relativistic effects up to second order in perturbation theory \cite{Matsubara:1995kq,Matarrese:1997ay,Boubekeur:2008kn,Bartolo:2010rw,Bruni:2013qta,Pajer:2013ana,Villa:2014foa}. Note that in the squeezed limit one needs to consider also the small scales which have entered the horizon during the radiation-dominated era. The early radiation effects have been explored in Refs.\ \cite{Fitzpatrick:2009ci,Tram:2016cpy} and the solution in the $\Lambda$CDM cosmology is discussed in Ref.\ \cite{Villa:2015ppa}. Intrinsic non-Gaussianities from relativistic effects, which are purely due to gravitational interaction in the radiation and matter dominated eras, are highly degenerate with PNG of the local type. Furthermore, they are of the same order as the constraints promised by upcoming surveys (see e.g.~\cite{DiDio:2016gpd}). It was shown in \cite{Castiblanco:2018qsd} that there are relativistic contributions to the bispectrum which have the same dependence in $k$-space (shape) and in redshift (time) as a primordial signal. These specific contributions come from setting adiabatic initial conditions, and account for projection effects in the galaxy bispectrum \cite{Kehagias:2015tda}. Furthermore, a complete discussion of galaxy bias in this context crucially depends on accounting for non-linear evolution in general relativity~\cite{Bartolo:2015qva}, and the correct initial conditions~\cite{Creminelli:2013mca, dePutter:2015vga} for the matter-dominated era.

Initial conditions for the evolution of structure in the late universe should account for the evolution of perturbations during radiation domination. Numerical codes solving the full Einstein-Boltzmann system in the $\Lambda$CDM cosmology up to second order were first developed in Refs.\ \cite{Pitrou:2008ut,Pettinari:2014vja}. In Ref.\ \cite{Tram:2016cpy}, the numerical code \texttt{SONG} \cite{Pettinari:2014vja} is used to compute the intrinsic matter bispectrum, which sets the initial conditions at second order for the subsequent evolution of structure. However, to correctly model the squeezed limit, one also needs to account for the small-scale physics. These scales eventually become highly non-linear so that a precise calculation needs to rely on simulations. In this paper we will use the relativistic N-body code \texttt{gevolution} \cite{Adamek:2015eda,Adamek:2016zes} which is based on a framework that is well adapted to our needs. In particular, all quantities are evolved in Poisson gauge that is also used in theoretical calculations.

In its basic implementation, \texttt{gevolution} assumes Gaussian initial conditions and only linear perturbations. In this work, however, we want to set initial conditions up to second order such that intrinsic non-Gaussianities from early gravitational evolution are accounted for. Indeed, for the purpose of computing the bispectrum, assuming Gaussian initial conditions at the start of a simulation would artificially impose a primordial non-Gaussian contribution that exactly cancels the intrinsic one. It was argued in Ref.\ \cite{Castiblanco:2018qsd} that higher-order corrections are subdominant in the initial conditions.

A generic non-Gaussian initial conditions generator has been implemented in Refs.\ \cite{Wagner:2010me,Wagner:2011wx} to study the scale-dependent halo bias. Further works have implemented non-Gaussian initial conditions thanks to the property of separability \cite{Scoccimarro:2011pz,Adhikari:2014xua,Smith:2010gx}. A separable modal decomposition of the bispectrum or trispectrum has also been worked out in Ref.\ \cite{Regan:2011zq}. This method was developed for the CMB bispectrum modal estimator \cite{Fergusson:2010dm} and was used in the \textit{Planck} analysis \cite{Akrami:2019izv}. The modal estimator has also been extended to studies of large-scale structure in Ref.\ \cite{Fergusson:2010ia} and more recently in Ref.\ \cite{Hung:2019ygc}. For this work, given that the kernel produced by \texttt{SONG} is not separable, we will use a method similar to the one of Refs.\ \cite{Wagner:2010me,Wagner:2011wx} as it can easily be extended to include the relativistic effects. We leave a modal approach extension for future work. Recent work \cite{Enriquez:2021arn} used the equivalence between the Newtonian Lagrangian frame and the relativistic comoving synchronous frame in order to incorporate some second- and third-order terms in the initial conditions of Newtonian N-body simulations (which are relativistic in the sense of said correspondence). This approach does not account for the whole second-order transfer function (e.g.\ the non-linear evolution of the plasma in the radiation dominated era).

In this paper, we present a generator of RELativistic second-order Initial Conditions (\texttt{RELIC}) for the N-body code \texttt{gevolution}. In Section \ref{sec:Second-order initial condition},  we lay down the basic equations of our second-order initial conditions and present our method, similar to Refs.\ \cite{Wagner:2010me,Wagner:2011wx} that keep the computational time of our second-order quantities at a manageable level. In Section \ref{sec:gevolution}, we describe how the second-order quantities that we compute can be used in the relativistic code \texttt{gevolution}. In Section \ref{sec:checks}, we perform several consistency checks at the level of the two-point and three-point correlation functions, as well as further checks that our second-order initial condition are well propagated in the N-body experiments. In Section \ref{sec:concl}, we conclude on this work and propose some follow-up avenues to continue the modelling and propagation of primordial non-Gaussianities.

\section{Second-order initial conditions}
\label{sec:Second-order initial condition}
The main goal of the RELativistic Initial Conditions generator \texttt{RELIC} is to compute the second-order initial Cauchy data for relativistic simulations of large-scale structure. We start by expanding perturbed fields $\I$ (such as the density contrast) in terms of the primordial comoving curvature perturbation $\R(\mathbf k)$. The field $\R(\mathbf k)$ is set when the mode $\mathbf{k}$ is far outside the horizon, e.g.\ at the end of inflation. At linear order the modes are decoupled in Fourier space \cite{Baumann:2009ds}, hence we can write $\I$ as
\begin{equation}
\label{eq:transfer_I}
\I  (\tau,\mathbf k) = \T^{(1)}_{\mathcal I}(\tau,k) \R (\mathbf k)\,,
\end{equation}
where we have defined the first-order \textit{transfer function} $\T^{(1)}_\mathcal{I}(\tau,k)$. The transfer function is a deterministic quantity that contains the linear evolution of the modes while $\R(\mathbf k)$ is the stochastic initial condition.

At second order modes couple. Since the quadratic terms turn into convolution integrals in Fourier space, one can define the second-order transfer function $\T^{(2)}_{\I}$ as their kernel so that \cite{Pettinari:2014vja, Pitrou:2010sn}: 
\begin{equation}
    \label{eq:transfer2}
    \I (\tau,k) \equiv \I^{(1)}+ \I^{(2)} =\T^{(1)}_{\I}(\tau,k) \R(\mathbf k) +  \int_{\mathbf {k_1}\mathbf { \mathbf {k_2}}}   \T^{(2)}_{\I}(\tau,k_1,k_2,k) \R (\mathbf k_1) \R (\mathbf k_2)  \,,
\end{equation}
where $\I^{(1)}$ is first order in perturbations and $\I^{(2)}$ is second order. The integral shorthand is defined as
\begin{equation}
    \label{eq:def_int}
    \int_{\mathbf {k_1 k_2}} = \int \frac{d^3 k_1 d^3k_2}{(2\pi)^3} \delta(\mathbf {k}-\mathbf {k_1}-\mathbf {k_2})\,.
\end{equation}
The second term of Eq.\ \eqref{eq:transfer2} contains the non-linear evolution of the modes with an explicit mode coupling. 

We can compute the power spectrum $P_\I(k)$ of the field $\I$ defined as
\begin{equation}
    \label{eq:powerspectrum_def2}
        \left< \I(\mathbf k_1) \I(\mathbf k_2) \right> = (2\pi)^3 \delta(\mathbf k_1+ \mathbf k_2) P_\I(k_2)\,,
\end{equation}
and the bispectrum $B_\I(k_1,k_2,k_3)$ defined as
\begin{equation}
    \label{eq:bispectrum_def}
        \left< \I(\mathbf k_1) \I(\mathbf k_2) \I(\mathbf k_3)\right> = (2\pi)^3 \delta(\mathbf k_1+ \mathbf k_2+ \mathbf k_2) B_\I(k_1,k_2,k_3)\,.
\end{equation}
We can also compute the tree-level bispectrum at any redshift by plugging \eqref{eq:transfer2} into Eq.\ \eqref{eq:bispectrum_def}. It reads \cite{Pettinari:2014vja}
\begin{multline}
    \label{eq:Bisp_decomp}
        B_{\I}(k_1,k_2,k_3) = \T^{(1)}_{\I}(k_1) \T^{(1)}_{\I}(k_2) \T^{(1)}_\I(k_3) B_{\R}(k_1,k_2,k_3)\\
                           +2 \T^{(1)}_{\I}(k_1) \T^{(1)}_{\I}(k_2) \T^{(2)}_{\I}(k_1,k_2,k_3) P_{\R}(k_1)P_{\R}(k_2)+ \text{2 perms.}\,.
\end{multline}
The first term on the right-hand side is the primordial bispectrum of the curvature perturbation. It vanishes if we assume Gaussian primordial initial conditions. The second term is called the intrinsic bispectrum and comes from the non-linear dynamics. As explained in Section \ref{sec:Introduction}, the second-order relativistic and early radiation effects contained in the second term are degenerate in time and in their scale dependence with the local type of PNG, see for example Ref.\ \cite{Castiblanco:2018qsd,Tram:2016cpy}. Hence, in order to measure accurately the PNG, one needs to know these contributions. 

\subsection{Efficient generation of second-order initial conditions with arbitrary kernels}
\label{sec:nGfield}

The computation of second-order quantities involves the convolution integral of Eq.\ \eqref{eq:transfer2}. Using the Dirac delta function to remove one integral, Eq.\ \eqref{eq:transfer2} can be rewritten as
\begin{equation}
    \label{eq:transfer22}
    \I^{(2)} (\tau ,\mathbf k) = \int \frac{d^3k_1}{(2\pi)^3} \T^{(2)}_{\I}(\tau,k_1,|\mathbf {k}-\mathbf {k_1}|,k) \R (\mathbf k_1) \R (\mathbf k - \mathbf k_1)\,.
\end{equation}
Hence, for a cubic grid of linear size $N$, the time complexity to compute the second-order fields is $N^6$ which can mean that the generation of the initial Cauchy data becomes the most expensive part of a simulation. For large $N$ such a brute-force approach becomes impractical.

Taking inspiration from effective theories, our proposal for decreasing the time complexity is to introduce a split between long and short scales and to consider only the long-long and long-short mode couplings, dismissing the short-short mode couplings. With a fixed number $N_\Lambda^3$ of long modes this will give a time complexity of $N_\Lambda^3 N^3$ which is much more manageable for large $N$. For that purpose, we introduce a scale cut $k_{\Lambda}$ that separates ``long'' and ``short'' scales. Knowing that relativistic effects are relevant only when large scales are involved, we split the curvature perturbation in two parts $\R \equiv \R_L+\R_S$ such that
\begin{equation}
  \R_{L}(k)=W(k)\R(k), \qquad \R_{S}(k)=(1-W(k))\R(k)\,,
\end{equation}
where $W(k)$ is a window function: $W(k)=1$ if $k<k_{\Lambda}$ and $W(k)=0$ otherwise. 
We replace $\R$ in Equation \eqref{eq:transfer22} by the sum $\R_L+\R_S$ to get:
\begin{equation}\label{int2}
\begin{split}
  \I^{(2)}(\tau ,\mathbf k)= &\int \frac{d^3 k_1}{(2 \pi)^3} \T^{(2)}_{\I} (\tau,k_1, |\mathbf {k}-\mathbf {k_1}|,k) \R_L(\mathbf{k_1})\R_L(\mathbf{\mathbf{k}-\mathbf{k_1}}) \\
+ &\int \frac{d^3 k_1}{(2 \pi)^3} \T^{(2)}_{\I} (\tau,k_1, |\mathbf {k}-\mathbf {k_1}|,k) \R_S(\mathbf{k_1})\R_S(\mathbf{\mathbf{k}-\mathbf{k_1}}) \\
    +2 &\int \frac{d^3 k_1}{(2 \pi)^3} \T^{(2)}_{\I} (\tau,k_1, |\mathbf {k}-\mathbf {k_1}|,k) \R_S(\mathbf{k_1})\R_L(\mathbf{\mathbf{k}-\mathbf{k_1}})\\
    &\equiv \I_{LL}+\I_{SS}+2\I_{LS}\,.
\end{split}
\end{equation}

The first term $\mathcal I_{LL}$ takes into account the coupling between large scales. This term was already considered in the method of Ref.\ \cite{Wagner:2011wx}. Relativistic corrections are important at large scales so that $\mathcal I_{LL}$ is relevant for us. Its time complexity scales like $N_{\Lambda}^6$.

The second term accounts for the coupling between large and small scales, i.e.\ a squeezed configuration in the bispectrum. The relativistic effects peak in this limit and are degenerate with local PNG, so it is crucial to compute it. Its time complexity is $N^3N_\Lambda^3$.

The last term $\mathcal I_{SS}$ contains the coupling between small scales. Its time complexity is $N^6$. For the purpose of PNG and relativistic effects, the coupling between small modes is dominated by Newtonian non-linearities which are negligible at the initial time since they scale like $a^2$. Note however that, like the other terms, $\mathcal I_{SS}(\mathbf k)$ contributes to the power at large scales $k>k_\Lambda$. If we neglect the raw $\mathcal I_{SS}$, we would miss this power and bias the large scales which we want to be accurately computed. To split the large- and small-scale contributions, we can again apply the window function:
\begin{equation}\label{eq:SSsplit}
    \mathcal I_{SS}(\mathbf {k})=W(k)\mathcal I_{SS}(\mathbf {k})+(1-W(k))\mathcal I_{SS}(\mathbf {k})=\mathcal I_{SS}^L(\mathbf{k})+\mathcal I_{SS}^S(\mathbf{k})\,.
\end{equation}
This way, the $\I_{SS}^L$ contains all the contributions of the small-small coupling to the large scales $k<k_{\Lambda}$. The number of operations per point is still $N^3$ but we only have to compute it for $k<k_\Lambda$ since it vanishes anywhere else. Its time complexity is therefore $N^3N_\Lambda^3$, i.e.\ like the squeezed term. Finally, the term $\I_{SS}^S(\mathbf{k})$ is the one that scales like $N^6$ and accounts for the small-small mode coupling for the small modes. This is the term that we neglect in our analysis.

Our final approximation takes the form 
\begin{equation}
    \label{eq:final_approx}
 \I^{(2)}(\mathbf k) \approx \mathcal I_{LL} (\mathbf {k})+ 2  \mathcal I_{SL} (\mathbf {k})+ \mathcal I_{SS}^L(\mathbf {k})\,.
\end{equation}
The $\I$'s are defined in Eqs.\ \eqref{int2} and \eqref{eq:SSsplit}. Note that Eq.\ \eqref{eq:final_approx} is exact at large scales $k<k_{\Lambda}$ and is therefore only an approximation in the range $k>k_{\Lambda}$.

\subsection{Relativistic initial conditions at second order}\label{sec:analytical}

We apply the method described in the previous section to the generation of second-order initial conditions. We wish these to be relativistic and to account for the non-linear evolution during radiation domination. Such initial conditions can be generated with the help of the code \texttt{SONG} that can compute the second-order transfer function of the density contrast of cold dark matter (CDM). We will then use them to run relativistic simulations using the relativistic N-body code \texttt{gevolution}.

We work in Poisson gauge with the following line element,
\begin{equation}
\label{eq:metric}
 ds^2 = -a^2 e^{2\psi} d\tau^2 + a^2 e^{-2\phi} \delta_{ij}  \left(dx^i + \beta^i d\tau\right)  \left(dx^j + \beta^j d\tau\right).
\end{equation}
Tensor modes can be consistently ignored at second order for the purpose of setting initial conditions.\footnote{We ignore linear tensor modes in the initial conditions. Tensor modes are sourced by the coupling of first-order fields at second order, but they will again decouple from other second-order perturbations. Thus, they need to be considered starting at third order.} The relevant gauge conditions are that the covariant divergence of $\beta^i$ on the spatial hypersurface vanishes. At second order, it is enough to require that $\sum_i \partial\beta^i/\partial x^i = 0$ because $\beta^i$ is already a second-order quantity and the covariantisation of this condition only adds higher-order terms. Our conventions are that a prime represents a derivative with respect to conformal time $\tau$: $' \equiv \frac{\partial}{\partial \tau}$, and the conformal Hubble factor is defined as the derivative of the logarithm of the scale factor with respect to the conformal time: $\mathcal{H} \equiv \frac{\partial \ln a}{\partial \tau}$.

The code \texttt{SONG}  provides the second-order transfer function for the density that is defined as following for a CDM perfect fluid,
\begin{equation}\label{delta_song}
T^{\mu\nu} = \bar{\rho} \left(1 + \delta_\mathrm{fl}\right) u^\mu u^\nu\,,
\end{equation}
where, defining $v^\mu = dx^\mu/d\tau$, we have $u^\mu = v^\mu / \sqrt{-v^\nu v_\nu}$. We note that the density contrast defined this way is not the one of Poisson gauge which is given by\footnote{It is worth pointing out that this definition does not require the stress-energy to be of any particular form, while Eq.\ \eqref{delta_song} only works for a perfect fluid.}
\begin{equation}\label{delta_gev}
 \rho \equiv n_\mu n_\nu T^{\mu\nu} = a^2 e^{2\psi} T^{00} = \bar{\rho} \left(1 + \delta\right)\,,
\end{equation}
where the vector $n_\mu$ is the unit normal on the equal-time hypersurface of the Poisson gauge. At first order the convention of \texttt{SONG} agrees with the Poisson gauge, but at second order some differences appear because the fluid four-velocity is not normal to the hypersurfaces. We get
\begin{equation}\label{eq:delta}
    \delta = \delta_\mathrm{fl} + v^2 + \ldots\,,
\end{equation}
where the ellipsis stands for higher-order terms we need not consider here. It is also sufficient to use the first-order velocity in this relation.

Since a CDM perfect fluid really only has one degree of freedom, once the density has been specified all other fields (velocity, metric perturbations) are determined by constraints. For later convenience we introduce the gravitational slip $\chi = \phi - \psi$. The first-order fields are related to the primordial curvature perturbation via the usual transfer functions that can be computed with a linear Boltzmann code. While we will assume matter domination, these first-order solutions may include small contributions from relativistic species (neutrinos and radiation). However, we will assume that such contributions to the second order can be neglected. This also means, for example, that we can use the fact that $\phi_{(1)}^2 \gg \mathcal{H}^2 (\phi'_{(1)})^2$ and neglect terms that are quadratic in the time derivatives of first-order potentials.

\texttt{RELIC} first generates realisations of all first-order fields (using transfer functions from \texttt{CLASS} \cite{Blas:2011rf}) and then uses the method described in the previous Section \ref{sec:nGfield} to generate a realisation of $\delta^{(2)}_\mathrm{fl}$ (using the respective second-order kernel from \texttt{SONG}). In this article, we take the second-order matter density contrast to be given only by the CDM contribution. We leave the inclusion of baryons at second order for future work. However, at first order, we use a weighted sum of the CDM and the baryon density and velocity fields. Neglecting the effect of baryons at first order would potentially generate an error larger than the size of the second-order terms we keep. Starting from the first-order fields and the solution for $\delta^{(2)}_\mathrm{fl}$ we now compute the remaining variables up to second order. Only the density, the canonical momentum and the potential $\phi$ will actually be required for setting the initial Cauchy data in the N-body code \texttt{gevolution} -- the remaining variables will be obtained automatically by solving the constraints within the code.

At second order, the gravitational slip is sourced by the matter anisotropic stress via the elliptic constraint
\begin{equation}
\label{eq:chi2}
    \Delta^2 \chi_{(2)} = \frac{3}{2} \left(\nabla_i \nabla_j - \frac{1}{3} \delta_{ij} \Delta\right) \left(2 \nabla^i \psi_{(1)} \nabla^j \psi_{(1)} - \nabla^i \chi_{(1)} \nabla^j \chi_{(1)} + 3 \mathcal{H}^2 \Omega_m v^i_{(1)} v^j_{(1)}\right)\,.
\end{equation}
We may drop the term quadratic in $\chi_{(1)}$ because $\chi_{(1)}$ is only sourced by the anisotropic stress of relativistic species. The spatial trace of Einstein's equations yields
\begin{equation}
    3 \phi''_{(2)} + 9 \mathcal{H} \phi'_{(2)} = \Delta \chi_{(2)} + 3 \mathcal{H} \chi'_{(2)} - \frac{1}{2} \left(\nabla \phi_{(1)}\right)^2 + \frac{3}{2} \mathcal{H}^2 \Omega_m v^2_{(1)}\,.
\end{equation}
In matter domination all the terms on the right-hand side are approximately constant, and with $\mathcal{H}' \simeq -\mathcal{H}^2 / 2$ the growing mode solution is given by
\begin{equation}
\label{eq:phiprime2}
    \phi'_{(2)} \simeq \frac{2 \Delta \chi_{(2)} - \left(\nabla \phi_{(1)}\right)^2 + 3 \mathcal{H}^2 \Omega_m v^2_{(1)}}{21 \mathcal{H}}\,.
\end{equation}
The Hamiltonian constraint finally gives an equation for $\phi_{(2)}$,
\begin{multline}
\label{eq:phi2}
    \Delta \phi_{(2)} - 3 \mathcal{H}^2 \phi_{(2)} =\\ 3 \mathcal{H} \phi'_{(2)} - 3 \mathcal{H}^2 \chi_{(2)} - 3 \mathcal{H}^2 \psi^2_{(1)} - 2 \phi_{(1)} \Delta \phi_{(1)} + \frac{1}{2} \left(\nabla\phi_{(1)}\right)^2 + \frac{3}{2} \mathcal{H}^2 \Omega_m \left(\delta^{(2)}_\mathrm{fl} + v^2_{(1)}\right)\,.
\end{multline}

At second order, the momentum constraint reads
\begin{multline}
    \frac{1}{4} \delta_{ij} \Delta \beta^j_{(2)} - \nabla_i \phi'_{(2)} - \mathcal{H} \nabla_i \psi_{(2)} 
    + \phi'_{(1)} \nabla_i \psi_{(1)} =\\ \frac{3}{2} \mathcal{H}^2 \Omega_m \delta_{ij} \left[v^j_{(2)} + \beta^j_{(2)} + v^j_{(1)} \left(\delta^{(1)}_\mathrm{fl} - 2 \phi_{(1)} 
    \right)\right]\,.
\end{multline}
Subtracting $\psi_{(1)}$ times the first-order constraint and taking the divergence yields following useful relation,
\begin{multline}
\label{eq:velocitydivergence}
    \Delta \left(\frac{1}{2}\mathcal{H} \psi_{(1)}^2 + \psi_{(1)} \phi'_{(1)} - \mathcal{H} \psi_{(2)} - \phi'_{(2)}\right) - \frac{3}{2} \mathcal{H}^2 \Omega_m \nabla_i \left(v^i_{(1)} \delta_\mathrm{fl}^{(1)}\right) =\\ \frac{3}{2} \mathcal{H}^2 \Omega_m \nabla_i \left(v^i_{(2)} - 2 v^i_{(1)} \phi_{(1)} - v^i_{(1)} \psi_{(1)}\right)\,.
\end{multline}
We could of course solve this directly for the divergence of $v^i_{(2)}$ but we will see shortly that the right-hand side is exactly the combination of terms that sets the canonical momentum at second order.

Let us finally consider the curl part of the second-order velocity. We shall follow the argument of Ref.\ \cite{Lu:2008ju} which essentially states that, for a CDM perfect fluid, vorticity cannot be generated by scalar perturbations. The invariant object to consider here is the vorticity tensor measured in the fluid frame, i.e.\ 
\begin{equation}
    \omega_{\mu\nu} = h^\alpha_{\mu} h^\beta_{\nu} \nabla_{[\beta} u_{\alpha]}\,,
\end{equation}
where the square brackets denote anti-symmetrization and $h^\mu_\nu = \delta^\mu_\nu + u^\mu u_\nu$ is the projection into the fluid frame. If no vorticity is present initially, the vorticity tensor $w_{\mu\nu}$ remains zero. At second order this implies
\begin{equation}
    \omega^{(2)}_{ij} = a \delta_{k[i} \left[\nabla_{j]} v^k_{(2)} + \nabla_{j]} \beta^k_{(2)} - 2 v^k_{(1)} \nabla_{j]} \left(\phi_{(1)} + \psi_{(1)}\right) + \delta_{j]l} \left(v_{(1)}^k\right)' v_{(1)}^l\right] = 0\,.
\end{equation}
The last term can be rewritten using the geodesic equation at first order, $(v^k_{(1)})' + \mathcal{H} v^k_{(1)} + \nabla^k \psi_{(1)} = 0$, such that
\begin{equation}
\label{eq:vorticity}
    \omega^{(2)}_{ij} = a \delta_{k[i} \nabla_{j]} \left[v^k_{(2)} + \beta^k_{(2)} - v^k_{(1)} \left(2 \phi_{(1)} + \psi_{(1)}\right)\right] = 0\,.
\end{equation}
From the momentum constraint we then get
\begin{equation}
    \frac{1}{4} \delta_{k[i} \nabla_{j]}\Delta \beta^k_{(2)} = \frac{3}{2} \mathcal{H}^2 \Omega_m \delta_{k[i} \nabla_{j]} \left(v^k_{(1)}\delta^{(1)}_\mathrm{fl}\right)\,.
\end{equation}
This result could again be used to get the curl part of the velocity from Eq.\ \eqref{eq:vorticity}. However, we will see shortly that Eq.\ \eqref{eq:vorticity} directly corresponds to the curl of the canonical momentum that is needed for particle initial data.

In summary, once the realisation of the non-Gaussian field $\delta_\mathrm{fl}$ has been generated, other second-order fields are successively computed by solving Eqs.\ \eqref{eq:chi2}, \eqref{eq:phiprime2}, \eqref{eq:phi2} and \eqref{eq:velocitydivergence}. The terms that are quadratic in first-order perturbations are evaluated in configuration space where the multiplications are local, and Fourier transforms are only carried out to invert the linear operators acting on the unknown fields. This means that once the difficult task of generating the non-Gaussian field $\delta_\mathrm{fl}$ has been accomplished, the remaining operations have a time complexity of $N^3 \log N$.

\section{Implementation in \texttt{gevolution}}
\label{sec:gevolution}

In this section we explain how the fields that describe perturbations up to second order are used to populate the initial particle phase-space in \texttt{gevolution}. The code uses canonical momenta as phase-space coordinates which are related to the peculiar velocities as
\begin{equation}
    q_i = a m_p \frac{e^{-2\phi} \delta_{ij} \left(v^j + \beta^j\right)}{\sqrt{e^{2\psi} - e^{-2\phi} \delta_{kl} \left(v^k + \beta^k\right) \left(v^l + \beta^l\right)}}\,,
\end{equation}
where $m_p$ is the rest mass of the particle. We will write $q_p^2 = \delta^{ij} q_i q_j$ (here the subscript $p$ labels the different particles and is not to be confused with a spacetime index). Since $\beta^i$ vanishes at first order, we find at second order that
\begin{equation}
    q_i = a m_p \delta_{ij} \left(v^j + \beta^j - 2 \phi v^j - \psi v^j + \ldots\right)\,.
\end{equation}
We see from Eq.\ \eqref{eq:vorticity} that the curl of the momentum field vanishes even at second order for a CDM perfect fluid and hence it is sufficient to provide the divergence field $\nabla_i (v^i - 2 \phi v^i - \psi v^i)$ up to second order according to Eq.\ \eqref{eq:velocitydivergence}. An inverse Laplacian of this field then corresponds to a ``momentum potential'' such that the canonical momenta are given by its gradient. The initial value at a given particle position is obtained by interpolation from a regular mesh.

For a classical point-particle, the energy density is written as
\begin{equation}
    \rho_p(\mathbf{x}) \equiv a^2 e^{2\psi} T^{00}_{p}(\mathbf{x}) = \delta^3(\mathbf{x}-\mathbf{x}_p) \frac{m_p}{a^3 e^{-3\phi}} \sqrt{1 + e^{2\phi} \frac{q^2_p}{m_p^2 a^2}}\,.
\end{equation}
The total energy density of the $N$-body ensemble is given by the sum over all particles. In a particle-mesh scheme, the coarse-grained density $\hat{\rho} = \bar{\rho} (1 + \hat{\delta})$ is computed via a particle-to-mesh projection. Operationally, the stress-energy tensor is convolved with a kernel, e.g.\ the cloud-in-cell (CIC) kernel that is used in \texttt{gevolution},
\begin{equation}
    w_\mathrm{CIC}(\mathbf{x}) = \prod_{i=1}^3 \wedge(x^i)\,,\qquad\qquad \wedge(x) = \begin{cases}
    \ell^{-1} \left(1 + \frac{x}{\ell}\right) & \mathrm{if}~-\ell \leq x < 0\,,\\
    \ell^{-1} \left(1 - \frac{x}{\ell}\right) & \mathrm{if}~0 \leq x \leq \ell\,,\\
    0 & \mathrm{otherwise}\,,
    \end{cases}
\end{equation}
where $\ell$ is the resolution of the mesh. Hence,
\begin{equation}
\label{eq:cgdensity}
    \hat{\rho}(\mathbf{x}_g) = \sum_p w_\mathrm{CIC}(\mathbf{x}_g-\mathbf{x}_p) \frac{m_p}{a^3 e^{-3\phi}} \sqrt{1 + e^{2\phi} \frac{q^2_p}{m_p^2 a^2}}\,.
\end{equation}

The problem of generating $N$-body initial data can now be formulated as follows: given some fields $\hat{\rho}$, $\phi$, $q^2_p / m^2_p$, and after fixing $m_p$ appropriately, but to the same value for all particles, what are initial particle positions $\mathbf{x}_p$ such that Eq.\ \eqref{eq:cgdensity} is satisfied for all mesh points $\mathbf{x}_g$? In standard initial condition generators that use linear theory, one would require the errors to be at most second order in perturbations. Here, of course, we need to require the errors to be at most of third order. Setting $m_p$ to the same value for all particles is a performance consideration for $N$-body codes. Evidently, without this requirement the problem of initial conditions would have a trivial solution, namely placing one particle onto each mesh point and fixing its mass to satisfy Eq.\ \eqref{eq:cgdensity}. We will not consider this solution here.

We now present a general algorithm to solve the problem for $m_p$ fixed and identical for all particles. The algorithm is iterative such that the error can be successively reduced to higher and higher order as long as perturbation theory converges. We start by laying down a regular particle distribution, i.e.\ a ``crystal'' which has particle positions $\mathbf{x}^{(0)}_p$. Traditionally the density perturbations are imprinted by adding a small displacement, computed e.g.\ from Lagrangian perturbation theory within the continuum limit. Here we want to work directly with perturbation theory in the discrete system which allows us to have control over discretization errors that could easily overwhelm any second-order signals that we want to study. For our perturbative analysis to work well we deliberately choose the positions $\mathbf{x}^{(0)}_p$ such that $w_\mathrm{CIC}$ is continuously differentiable at all possible values $\mathbf{x}_g-\mathbf{x}_p^{(0)}$ --- this can be achieved by placing the particles away from the boundaries of the mesh cells, e.g.\ one particle at each center between eight nearest mesh points. To perturb the density we displace each particle by a small distance $\delta\mathbf{x}_p^{(1)}$ from its starting location. Taylor expanding Eq.\ \eqref{eq:cgdensity} we find at first order
\begin{equation}
    \hat{\rho}(\mathbf{x}_g) = \sum_p \left[w_\mathrm{CIC}(\mathbf{x}_g-\mathbf{x}_p^{(0)}) - \delta\mathbf{x}_p^{(1)} \cdot \nabla w_\mathrm{CIC}(\mathbf{x}_g-\mathbf{x}_p^{(0)}) + \ldots\right] \frac{m_p}{a^3 e^{-3\phi}} \sqrt{1 + e^{2\phi} \frac{q^2_p}{m_p^2 a^2}}\,.
\end{equation}
Moving the known quantities to the left and considering $\phi$ and $q_p$ as perturbative quantities we obtain
\begin{equation}
\label{eq:displacement1}
    \hat{\rho}(\mathbf{x}_g) - \bar{\rho} \left[1 + 3 \phi (\mathbf{x}_g)\right] = -\sum_p \delta\mathbf{x}_p^{(1)} \cdot \nabla w_\mathrm{CIC}(\mathbf{x}_g-\mathbf{x}_p^{(0)}) \frac{m_p}{a^3} + \ldots\,.
\end{equation}
This can be interpreted as a linear matrix equation with a known $N^3$-dimensional vector on the left-hand side, where $N$ is the number of mesh points in each space dimension, and a $3 N_p$-dimensional unknown vector on the right-hand side multiplying a $3 N_p \times N^3$-matrix with known entries, where $N_p$ is the total number of particles. Depending on the value of $N_p$ this system may be overdetermined or underdetermined.

The next step is to convert this equation into a solvable system. This is achieved by writing the $\delta\mathbf{x}_p^{(1)}$ as some discrete gradient of a \textit{displacement potential} $\xi^{(1)}$ that is defined on the mesh,
\begin{equation}
    \delta\mathbf{x}_p^{(1)} = \sum_{\mathbf{x}'_g} \mathbf{w}_\mathrm{grad}(\mathbf{x}'_g - \mathbf{x}_p^{(0)}) \xi^{(1)}(\mathbf{x}'_g)\,,
\end{equation}
where $\mathbf{w}_\mathrm{grad}$ is a known three-vector valued weight function that depends on the type of discrete gradient and interpolation method. Inserting this ansatz back into Eq.\ \eqref{eq:displacement1} one can now carry out the sum over particles and ends up with a new matrix equation for an unknown $N^3$-dimensional vector $\xi^{(1)}(\mathbf{x}'_g)$ that is multiplied by a known $N^3 \times N^3$ matrix. Furthermore, if we ensure that the ``crystal'' of particle positions $\mathbf{x}^{(0)}_p$ has at least the same discrete symmetries as the mesh, it is easy to see that the matrix coefficients are functions of $\mathbf{x}_g - \mathbf{x}'_g$ only, i.e.\ the right-hand side is a discrete convolution. The system can then be solved efficiently using fast Fourier transforms and the discrete convolution theorem.

Note that the existence of solutions generally depends on the choice of $\mathbf{w}_\mathrm{grad}$ and the initial particle positions $\mathbf{x}^{(0)}_p$. For some choices the convolution kernel becomes zero or very small for certain modes on the Fourier mesh, and these modes can then not, or not easily, be generated by the displacement in the way it is computed. Good results are obtained e.g.\ with a low-order gradient (we use a one-sided, two-point gradient) and low-order interpolation (we use a hybrid scheme that is a mix between nearest-grid-point and cloud-in-cell, as explained in appendix B of Ref.\ \cite{Adamek:2016zes}), both with one or with eight particles per mesh cell.

After this procedure we now have some first-order displacements $\delta\mathbf{x}_p^{(1)}$. We also have the full particle momenta which we assume can be interpolated from a mesh (the interpolation error does not matter since the momentum only contributes at order $q_p^2$ to the density) as well as the potential $\phi$. We can therefore compute a density $\hat{\rho}^{(1)}$ from the particle-to-mesh projection as
\begin{equation}
    \hat{\rho}^{(1)}(\mathbf{x}_g) = \sum_p w_\mathrm{CIC}(\mathbf{x}_g-\mathbf{x}_p^{(0)}-\delta\mathbf{x}_p^{(1)}) \frac{m_p}{a^3 e^{-3\phi}} \sqrt{1 + e^{2\phi} \frac{q^2_p}{m_p^2 a^2}}\,.
\end{equation}
Note that $\hat{\rho}^{(1)}$ is not defined as a term in a series expansion but as the result of the actual projection given the particle positions $\mathbf{x}_p^{(0)}+\delta\mathbf{x}_p^{(1)}$. It therefore contains various second and higher order contributions in $\delta\mathbf{x}_p^{(1)}$. However, it is clear that the residual $\hat{\rho} - \hat{\rho}^{(1)}$ is at least of second order in perturbations.

To improve our particle displacement, all we now need to do is to iterate the procedure once. That is, we write $\mathbf{x}_p = \mathbf{x}_p^{(0)}+\delta\mathbf{x}_p^{(1)}+\delta\mathbf{x}_p^{(2)}$ and expand Eq.\ \eqref{eq:cgdensity} for small $\delta\mathbf{x}_p^{(2)}$, treating $\delta\mathbf{x}_p^{(1)}$ as known quantity. We get
\begin{equation}
    \hat{\rho}(\mathbf{x}_g) - \hat{\rho}^{(1)}(\mathbf{x}_g) = -\sum_p \delta\mathbf{x}_p^{(2)} \cdot \nabla w_\mathrm{CIC}(\mathbf{x}_g-\mathbf{x}_p^{(0)}) \frac{m_p}{a^3} + \ldots\,,
\end{equation}
where only the right-hand side has also been expanded for small $\delta\mathbf{x}_p^{(1)}$ and resulting terms of order $\delta\mathbf{x}_p^{(1)} \delta\mathbf{x}_p^{(2)}$, formally third order, were relegated to the ellipsis. The computation of $\delta\mathbf{x}_p^{(2)}$ now follows the same procedure as was used for $\delta\mathbf{x}_p^{(1)}$ from Eq.\ \eqref{eq:displacement1} onward, just with a different input vector on the left-hand side. We arrive at corrected displacements that give rise to a projected density field $\hat{\rho}^{(2)}$ that coincides with the desired input field $\hat{\rho}$ up to third-order errors. The method can be iterated to further reduce the error to arbitrary order in a fashion that is reminiscent of a root-finding algorithm.

\section{Consistency checks}
\label{sec:checks}

In this section, we perform several checks on our initial conditions generator. First, we check the time complexity of \texttt{RELIC}. Then, we ensure that the code \texttt{RELIC} produces the expected matter fields from \texttt{SONG}. We do this by visual inspection and by measuring the power spectrum and bispectrum of the CDM density contrast $\delta_\mathrm{fl}$. Then, we pass the density, the canonical momentum and the potential fields to \texttt{gevolution} which  populates the initial particle phase-space. From this initial particle distribution, we verify that the particle-mesh projection reproduces the density field computed by \texttt{RELIC}. This provides a validation of the algorithm described in Section \ref{sec:gevolution}. Finally, in order to check the consistency of our pipeline, we evolve the simulation from redshift $z=100$ to $z=50$ and verify that the density field at $z=50$ computed by \texttt{gevolution} agrees with the density computed by \texttt{RELIC} at $z=50$.

For all these checks, we compute the initial realisations at redshift $z=100$ on a $N^3 = 512^3$ grid. The box size $L$ is chosen close to the cosmological horizon scale in order to be in the relativistic regime. We choose $L=3927~h^{-1}$\,Mpc which corresponds to a fundamental mode $k_{\text{min}}=1.6\times 10^{-3}~h$ Mpc$^{-1}$. 
Following the approximation presented in Section \ref{sec:nGfield}, we impose a scale cut at $k_{\Lambda}=4 \times 10^{-2}~h$\,Mpc$^{-1}$, i.e.\ $k_{\Lambda}=25 k_{\text{min}}$ and hence $N_\Lambda=25$. The cosmological parameters that we use are $h = 0.67556$ and otherwise taken as the best-fit values from Table 2 of Ref.\ \cite{Planck:2013pxb}, right column. We also assume three massless neutrino species.

In order to run \texttt{RELIC}, one needs first to compute the second-order transfer function of CDM. Here we use the second-order Boltzmann code \texttt{SONG}. The default setting of \texttt{RELIC} uses $k_{\text{min}}^{\texttt{SONG}}=k_{\text{min}}$.
We take $k_{\text{max}}^{\texttt{SONG}} = 0.9 N k_{\text{min}}$ close to the maximum modulus in our Fourier grid, $ \sqrt{3} N k_{\text{min}}/2$.
Since we use a linear sampling, in order to capture the large-scale features of the  second-order density transfer function, we use a step $dk = k_{\text{min}}$ so that $N^{\texttt{SONG}} =k_{\text{max}}^{\texttt{SONG}}/k_{\text{min}} $. This fixes the sampling of the first two modes of $\mathcal T^{(2)}(k_1,k_2,k_3)$. For the sampling of $k_3$ we use the ``smart'' configuration of \texttt{SONG}, see Ref.\ \cite{Pettinari:2014vja} for more details.

\subsection{Performance}%
\label{sub:Performance}

\begin{figure}\centering
\includegraphics[scale=1]{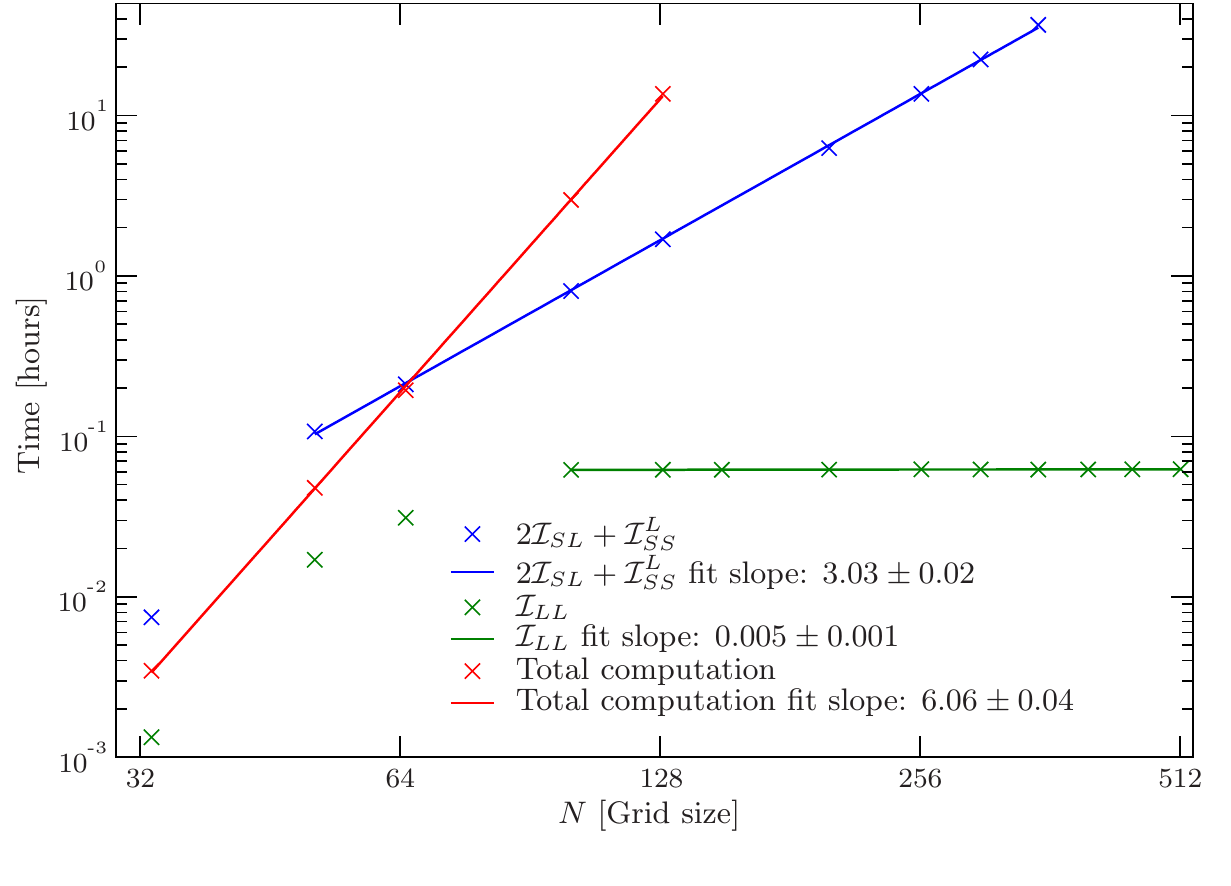}
    \caption{Computation time in hours as a function of the grid size $N$ in log-scale. The crosses are the computational time measured for each case and the solid line shows the result of the linear fit whose slope is indicated in the legend. We show separately the last two terms (in blue) and the first term (in green) of Eq.\ \eqref{eq:final_approx} since they have a different scaling. In red we show the time scaling of the full integral \eqref{eq:transfer22} without scale cut. The computation has been performed on 4 cores.}\label{fig:complexity}
\end{figure}

In Section \ref{sec:nGfield}, we used physical arguments to decrease the time complexity from $N^6$ to $N_\Lambda^3N^3$ thanks to the approximations on Eq.\ \eqref{eq:final_approx}. We present in Fig.~\ref{fig:complexity} the computational time on four cores as a function of the grid size $N$, keeping $N_\Lambda$ fixed. 
The \texttt{RELIC} approximation given in Eq.\ \eqref{eq:final_approx} is composed of three terms. The first one, $\mathcal I_{LL}$, is expected to have a constant time complexity. This is confirmed by the green curve in Fig.\ \ref{fig:complexity}. The blue curve contains the last two terms of Eq.\ \eqref{eq:final_approx}. We find that the code scales as expected. On $128$ cores similar to the ones used, initial conditions can be generated on a $1024^3$ grid in about a day. We also show in red the time scaling of the full integral \eqref{eq:transfer22}. As expected, it scales as $N^6$ with the grid size. On $128$ cores similar to the ones used, generating initial conditions on a $1024^3$ grid with the full integral would take more than ten years.

\subsection{Internal checks of \texttt{RELIC}}\label{sub:Internal checks}

We first check that the power spectra and bispectra of the fields generated by \texttt{RELIC} are indeed the ones provided by \texttt{CLASS} and \texttt{SONG}. We do this by measuring the power spectrum and bispectrum of the CDM density contrast on $20$ realisations. Relativistic effects are small (of the same order as the effects of $f_{\mathrm{NL}} \sim 1$). For this reason, we boost them by multiplying the second-order density contrast by a factor of $10^2$.

\subsubsection{Power spectra}

\begin{figure}[t]
\begin{center}
    \includegraphics[scale=1.3]{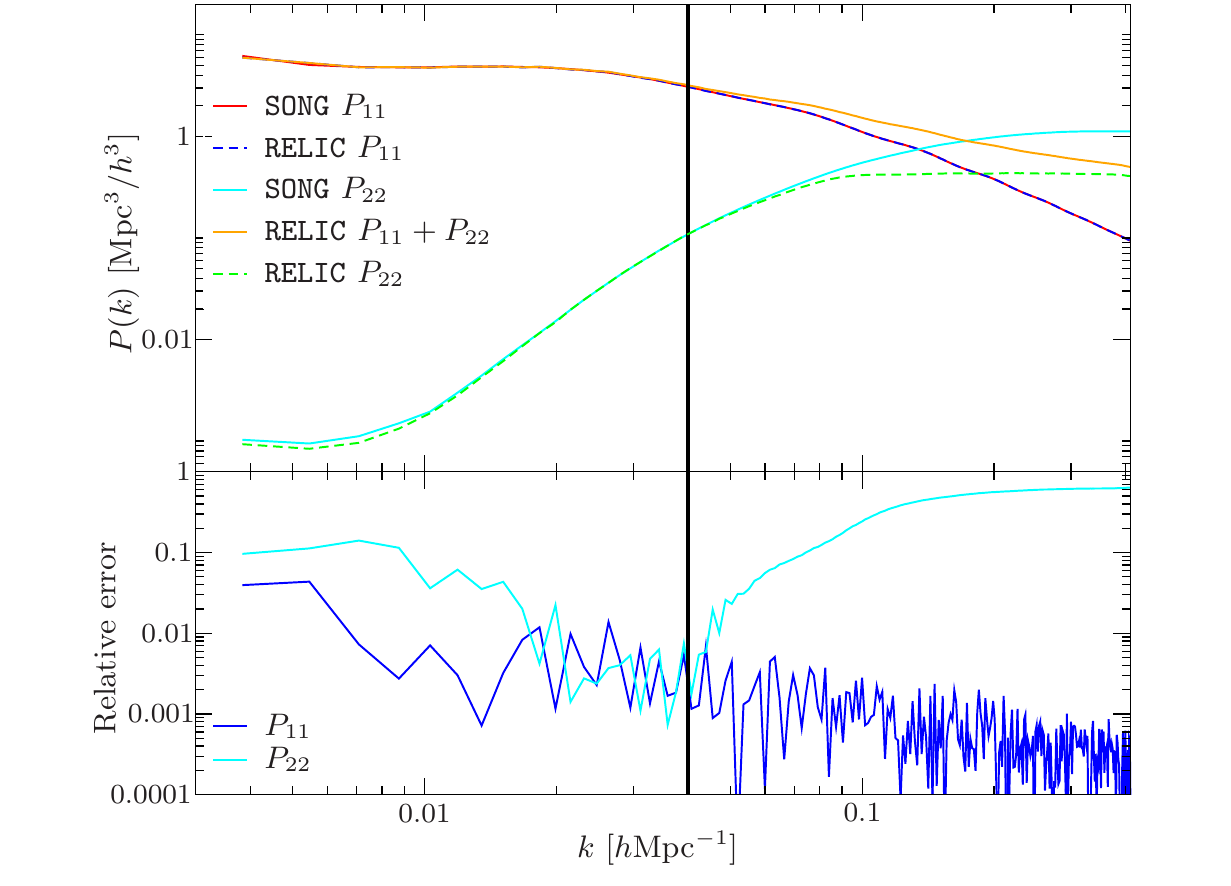}
\end{center}
\caption{ In the upper panel, the total power spectrum is shown in orange, i.e.\ the power spectrum of $\delta^{(1)}_\mathrm{fl}+10^2\delta^{(2)}_\mathrm{fl}$. The power spectra of the first- and the second-order component are shown in dashed blue and cyan, respectively. The red and green solid lines are the fiducial power spectra computed from Eqs.\ \eqref{eq:P_d1} and \eqref{eq:P_d2}. In the lower panel, we plot the relative difference between the fiducial spectra and the measurements (averaged over 20 realisations). The black vertical line indicates the cut-off scale $k_{\Lambda}=4\times10^{-2}~h$\,Mpc$^{-1}$.}
\label{fig:P_RIC}
\end{figure}

From the realisations of the first- and second-order density fields, $\delta^{(1)}_\mathrm{fl}$ and $\delta^{(2)}_\mathrm{fl}$, we measure their power spectra to check whether we recover the respective results from perturbation theory, $P_{11}$ and $P_{22}$ as computed from the transfer functions provided by \texttt{CLASS} and \texttt{SONG}.
In Fig.\ \ref{fig:P_RIC}, we plot the power spectrum of the field generated by \texttt{RELIC}, $\delta^{(1)}_\mathrm{fl}+10^2\delta^{(2)}_\mathrm{fl}$, in orange. In blue and cyan, we show the power spectra computed separately for $\delta^{(1)}_\mathrm{fl}$ (i.e.\ $P_{11}$) and for $10^2\delta^{(2)}_\mathrm{fl}$ (i.e.\ $10^4 P_{22}$). 
In red, we plot the first-order power spectrum given by
\begin{equation}
\label{eq:P_d1}
     P_{11}(k) = \left(\T^{(1)}_{\delta_{\mathrm{fl}}}(k)\right)^2 P_\R(k) \,,
\end{equation}
where the first-order transfer function is computed with \texttt{CLASS}. In green, we show the second-order density field power spectrum
\begin{equation}
\label{eq:P_d2}
     P_{22}(k) =  4\pi \int \frac{ q_1^2\sin{\theta}d q_1d\theta}{(2\pi)^3} \left(\mathcal T^{(2)}_{\delta_{\mathrm{fl}}}(q_1,k,\theta)\right)^2 P_{\mathcal R }(q_1) P_{\mathcal R}(q_1,k,\theta) 
    \,,
\end{equation}
where the second-order CDM transfer function is computed with \texttt{SONG}. The black vertical line indicates the cut-off scale $k_{\Lambda}$. We do not show the error bars since they are too small to be seen. Instead, we plot the relative error of $P_{11}$ and $P_{22}$, respectively, in blue and in cyan.

At first order, \texttt{RELIC} is at $0.1$ percent agreement with \texttt{CLASS} except for the largest scales, where there is a percent discrepancy. This is due to cosmic variance which induces larger variability at large scales. Indeed, the theoretical prediction is compatible with the scatter among realisations at large scales. Note also that the choice of binning is important since, at large scales, the number of modes available to measure the power spectrum becomes small. The tool we use, \texttt{Pylians3}, provides a routine that samples the fiducial power spectrum in the same way as the power spectrum estimated from the simulations. This effect depends on the exact scaling of the power spectrum for small $k$.

For the second-order power spectrum $P_{22}$, in cyan, we have a percent level agreement with the prediction from \texttt{SONG} in a range between $k=0.01~h$\,Mpc$^{-1}$ and the cut-off scale $k_{\Lambda}=4\times 10^{-2}~h$\,Mpc$^{-1}$. As expected, when $k>k_{\Lambda}$, our approximation breaks down and the measured power spectrum diverges from the fiducial one. At large scales the discrepancy is of order $10$\%. This is because the quadratic kernels in Eq.~\eqref{eq:P_d2} blow up in the infrared. Thus, the integral is very sensitive to the limits of integration, and there is a divergence that can be problematic for standard numerical integration routines (see e.g.\ Ref.\ \cite{Lewandowski:2017kes}). Isolating the divergences is non-trivial since the quadratic kernel was numerically computed from \texttt{SONG}. We instead rely on the bispectrum comparison of the next subsection for a more straightforward check.

\subsubsection{Bispectra}
\begin{figure}[t]
\begin{center}
    \includegraphics[scale=1.2]{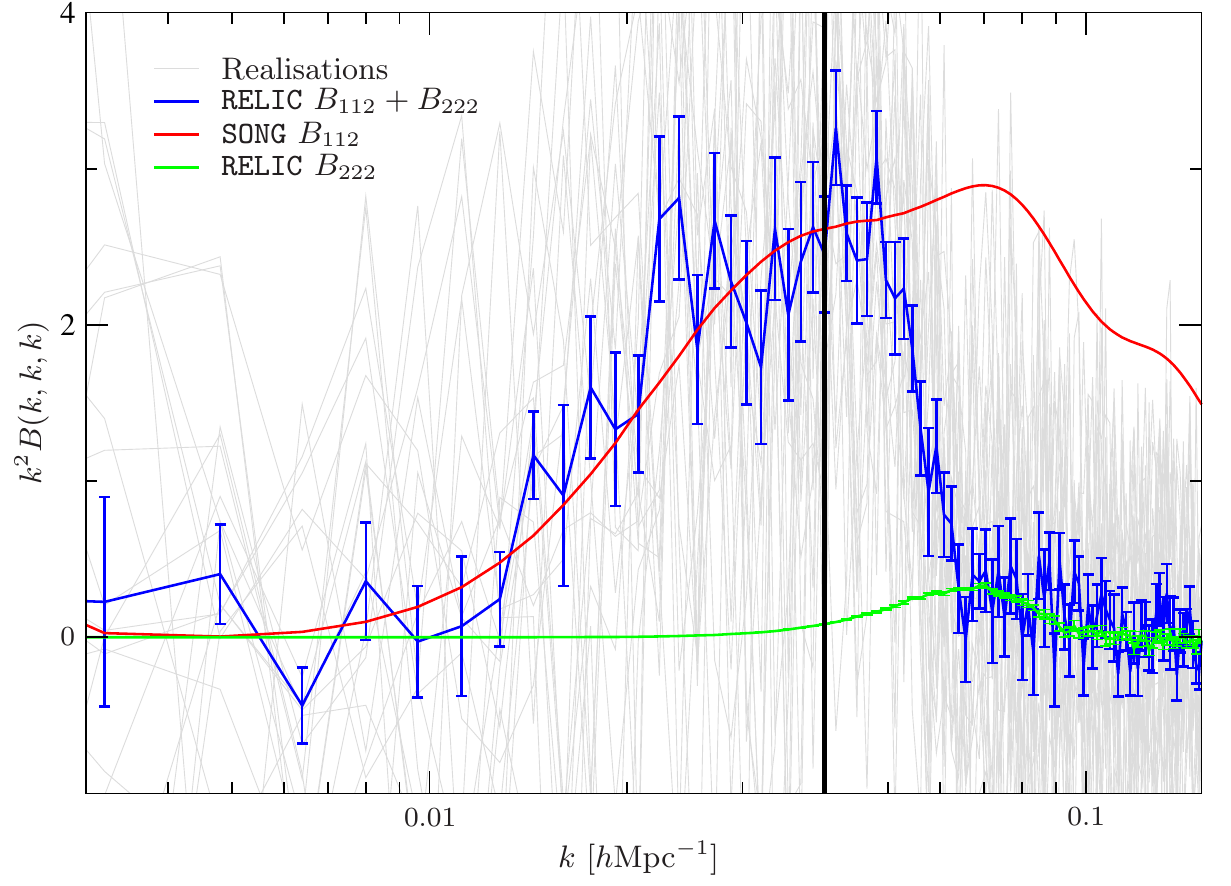}
\end{center}
    \caption{Equilateral configuration of the CDM density bispectrum. In red we plot the fiducial result computed from Eq.~\eqref{eq:Bisp_decomp} using \texttt{SONG}. The blue curve shows the measured average bispectrum from \texttt{RELIC}, which corresponds to $B_{112}+B_{222}$. The $B_{222}$ contribution is the bispectrum of $10^{2}\delta^{(2)}_\mathrm{fl}$, plotted in green. The vertical black line marks the cut-off scale $k_{\Lambda}=4\times10^{-2}~h$\,Mpc$^{-1}$.}
\label{fig:B_RIC}
\end{figure}

We measure the bispectrum of the initial density field generated by \texttt{RELIC} and compare it to the bispectrum given by the second-order kernel from \texttt{SONG} which was used as the input. If our numerical method works as expected, they should agree in the range of scales where the approximation of Section \ref{sec:analytical} holds.

In Fig.~\ref{fig:B_RIC} we plot the equilateral configuration of the bispectrum. In red, we show the fiducial bispectrum computed from Eq.~\eqref{eq:Bisp_decomp} using \texttt{SONG}.
In terms of perturbation theory, this corresponds to $B_{112}$.
In blue, we show the measurement of the \texttt{RELIC} realisations $\delta^{(1)}_\mathrm{fl} + 10^2 \delta^{(2)}_\mathrm{fl}$ which is $B_{112}+B_{222}$. Here $B_{222}$ is the bispectrum of the quadratic field $10^2\delta^{(2)}_\mathrm{fl}$, and corresponds to a one-loop contribution in standard perturbation theory. 
In green, we show $B_{222}$. Since we have boosted the second-order density, this contribution is boosted by a factor of $10^6$. We see that it remains negligible for the modes smaller than the cut-off. Our initial conditions generator reproduces the theoretical bispectrum as long as we consider modes smaller than the cut-off, indicated by the black vertical line. For modes larger than $3k_{\Lambda}$, the bispectrum falls to zero. 

\begin{figure}[t]
\begin{center}
    \includegraphics[scale=1.25]{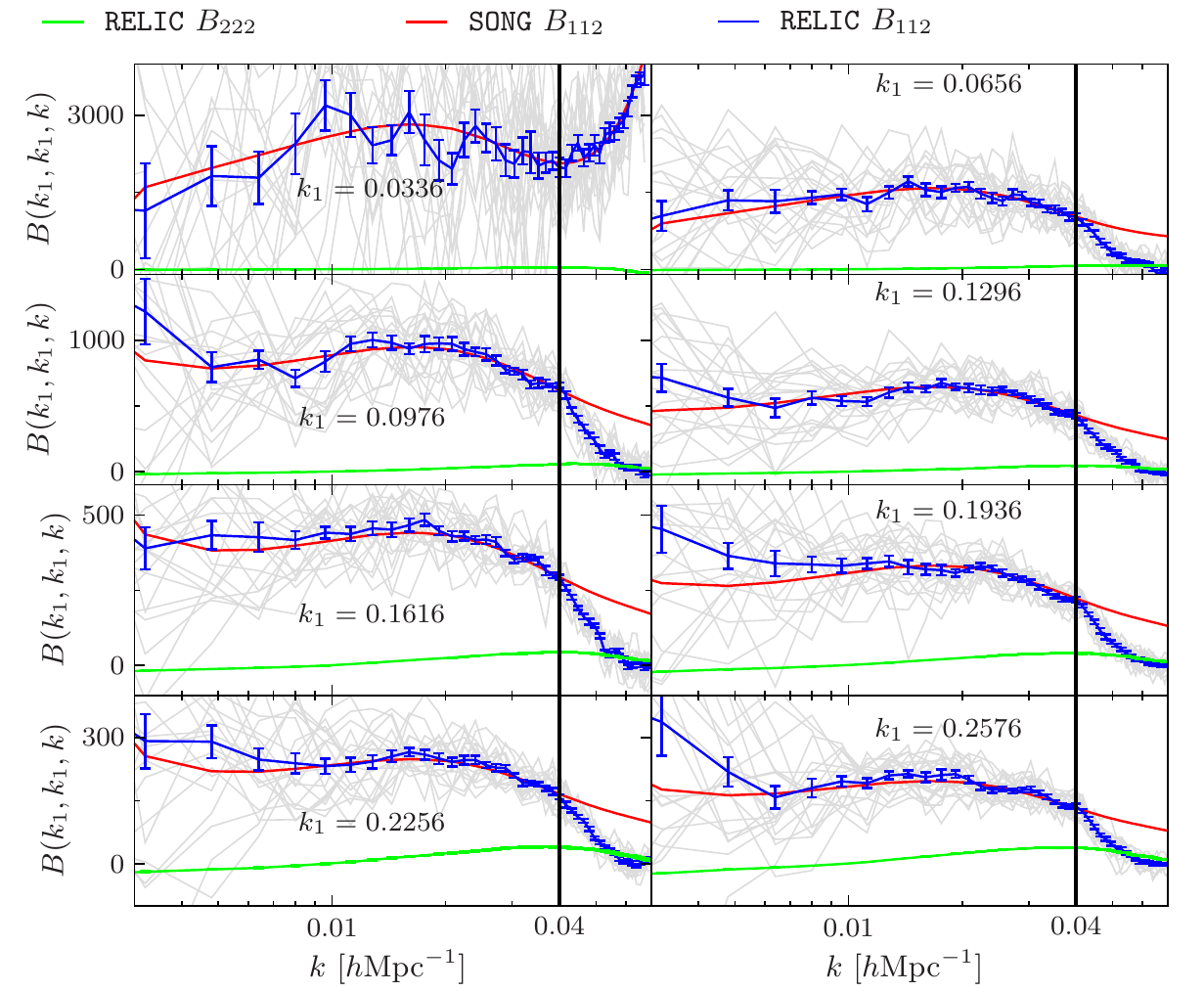}
\end{center}
    \caption{We plot for 8 squeezed configurations the CDM density bispectrum. The first two arguments are fixed to $k_1$ whose value in units of $h$\,Mpc$^{-1}$ is indicated above each panel. We vary the third argument between $k_\mathrm{min}$ and $2\times k_1$ to ensure the triangle inequality. The squeezed limit is obtained by taking $k$ much smaller than $k_1$. The red curve is the \texttt{SONG} bispectrum computed  from Equation \eqref{eq:Bisp_decomp}. The blue curve is the bispectrum of the \texttt{RELIC} realisations where we have subtracted the bispectrum of $10^2\delta^{(2)}$ plotted in green. The black vertical lines indicate the location of the cut-off scale $k_{\Lambda}=4\times 10^{-2}~h$\,Mpc$^{-1}$ in each panel.}
\label{fig:B_RIC_s}
\end{figure}

In Fig.~\ref{fig:B_RIC_s} we show the squeezed limit configuration of the CDM density bispectrum. Similarly to Fig.~\ref{fig:B_RIC}, the red and green curves are the fiducial bispectrum and the bispectrum of the quadratic field $B_{222}$. For each panel, we have fixed the first two arguments of the bispectrum to a given value, called $k_1$. The third argument varies between $k_\mathrm{min}$ and $2\times k_1$ according to the triangle inequality. The smaller $k$ is compared to $k_1$, the more the configuration is squeezed. For large $k_1$, the contribution of $B_{222}$ becomes important and biases the bispectrum of $\delta^{(1)}_\mathrm{fl}+10^2\delta^{(2)}_\mathrm{fl}$. Given the size of the error bars, this makes the measurement of the total bispectrum from \texttt{RELIC} incompatible with the fiducial bispectrum for the range $k\in[0.03,0.04]$. Hence, unlike in Fig.~\ref{fig:B_RIC}, we have subtracted the green curve $B_{222}$ from the total bispectrum, obtaining an agreement between the \texttt{RELIC} output and the \texttt{SONG} input.

The first panel corresponds to the lowest value of $k_1$ and here $k_1$ is smaller than the cut-off $k_{\Lambda}$. Our approximation holds as long as one of the modes in the bispectrum is smaller than $k_{\Lambda}$. Therefore, in this case the \texttt{RELIC} initial condition is accurate even for $k>k_{\Lambda}$. For all the other panels of Fig.~\ref{fig:B_RIC_s}, we have $k_1>k_{\Lambda}$ so 
that the \texttt{SONG} input and the \texttt{RELIC} output only agree for $k<k_{\Lambda}$. For smaller scales, the \texttt{RELIC} bispectrum falls to zero. 

Except for the first panel, we observe an overestimation of the first three  points, particularly visible in the sixth panel. This might be due to the choice of binning. The library \texttt{Pylians3} does not provide a binning routine for the fiducial bispectrum. Moreover, the sampling at large scales of \texttt{SONG} could also be improved (remember that we use a linear sampling with a step $dk=k_\mathrm{min}$).

\subsection{Tests of the implementation in \texttt{gevolution}}
\label{sec:Implementationgev}

\begin{figure}[tb]
\begin{center}
    \includegraphics[width=0.9\textwidth]{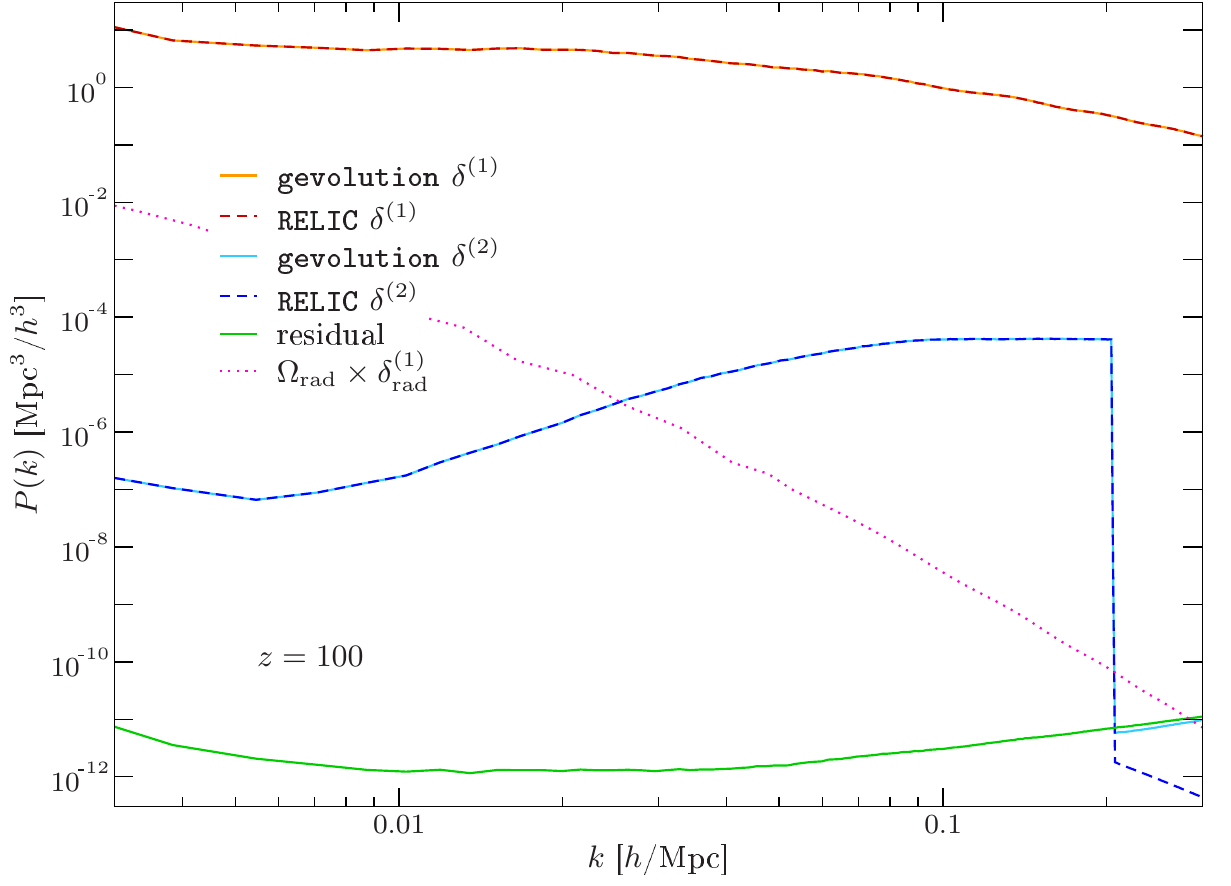}
\end{center}
    \caption{Power spectra of the initial density contrast at redshift $z=100$ in \texttt{RELIC} and in \texttt{gevolution}. The first-order density contrast in \texttt{gevolution} (computed with the first-order particle displacements only) is shown in orange and agrees well with the corresponding power spectrum of \texttt{RELIC} shown in dashed red. The second-order density contrast in the particle-mesh scheme (computed with the full particle displacements), shown as light blue curve, is obtained by subtracting the first order (\texttt{RELIC}) from the result of the particle-mesh projection. It, too, agrees very well with the second-order density contrast provided by \texttt{RELIC}, shown here as dashed blue curve. The residual, i.e.\ the difference between the full input field and the result of the particle-mesh projection, is shown in green. The sharp cut-off seen in the second order comes about because we choose to limit the respective calculation to wavenumbers smaller than $25\%$ of the Nyquist wavenumber. For comparison we also plot the first-order perturbations in the radiation field (dotted pink line).}
\label{fig:gevolution100}
\end{figure}

\begin{figure}[tb]
\centering
\begin{subfigure}{0.48\textwidth}
\includegraphics[width=\textwidth]{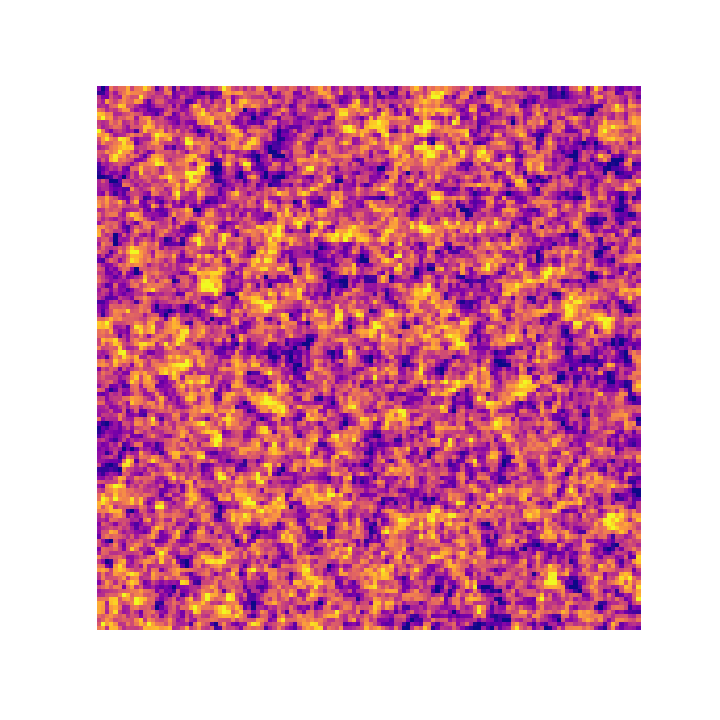}
\vspace{-1.5cm}
\caption{\texttt{RELIC} $\delta^{(1)}$}
\end{subfigure}
\hfill
\begin{subfigure}{0.48\textwidth}
\includegraphics[width=\textwidth]{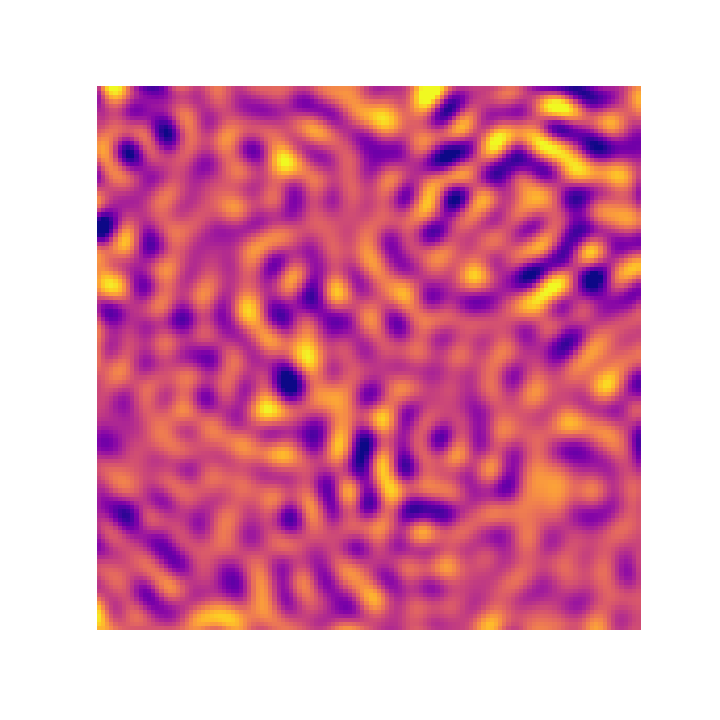}
\vspace{-1.5cm}
\caption{\texttt{RELIC} $\delta^{(2)}\times~200$}
\end{subfigure}
\hfill
\begin{subfigure}{0.48\textwidth}
\includegraphics[width=\textwidth]{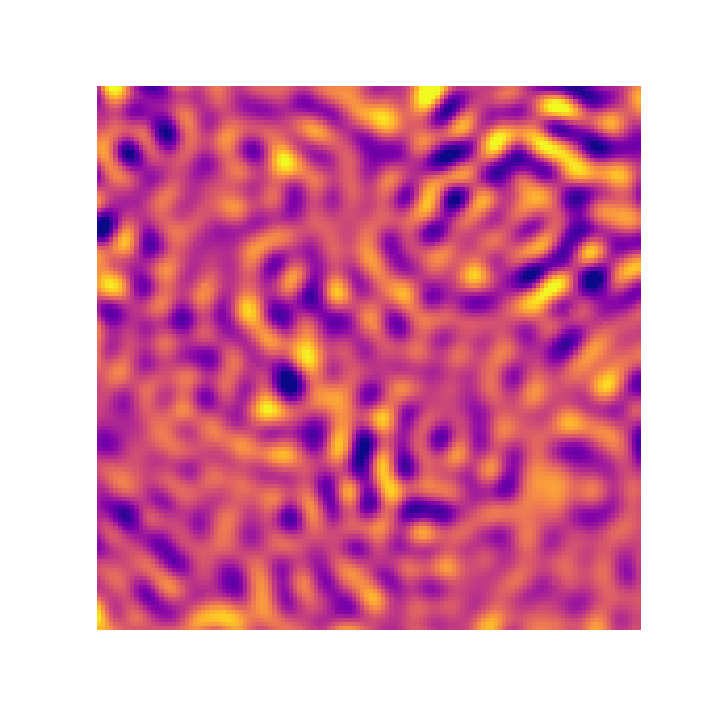}
\vspace{-1.5cm}
\caption{\texttt{gevolution} $\delta^{(2)}\times~200$}
\end{subfigure}
\hfill
\begin{subfigure}{0.48\textwidth}
\includegraphics[width=\textwidth]{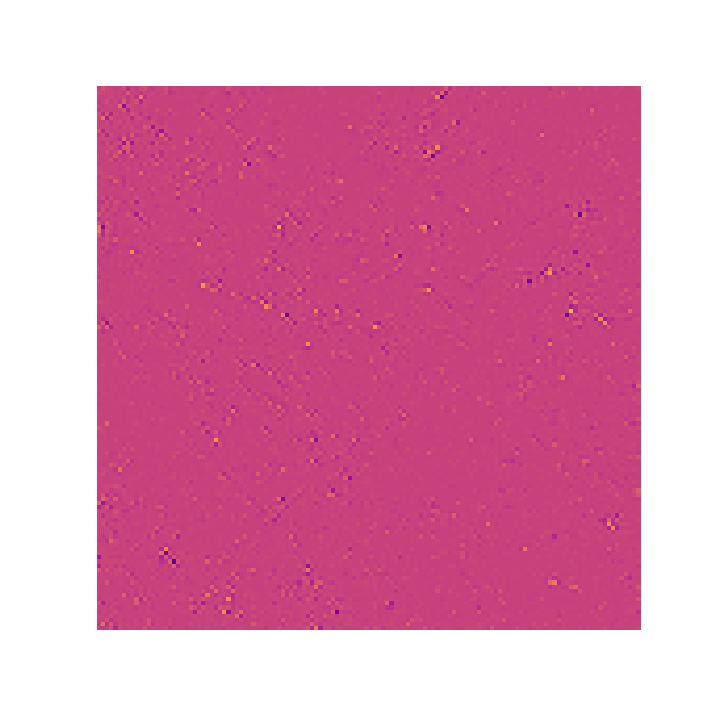}
\vspace{-1.5cm}
\caption{residual $\times~2000$}
\end{subfigure}
\caption{The four panels show a slice through a region of $500\,h^{-1}$\,Mpc at redshift $z=100$ -- the colour scale ranges $\pm 5 \%$ and is identical for all panels. Panel (a) shows the first-order density perturbation generated with \texttt{RELIC}. Panel (b) shows the second-order density perturbation generated with \texttt{RELIC}, multiplied by a factor $200$ to adjust the colour scale. While the full resolution is about $4\,h^{-1}$\,Mpc, a cut-off for computing $\delta^{(2)}$ was chosen close to $80\,h^{-1}$\,Mpc which explains the lack of features on smaller scales. Panel (c) shows the density contrast from the particle-to-mesh projection in \texttt{gevolution} with the first-order density perturbation of panel (a) subtracted, and multiplied by a factor $200$ like in panel (b). Panel (d) finally shows the difference between panels (c) and (b), multiplied by another factor of $10$ for a total factor of $2000$, and represents the residual between the input density contrast (the sum of first and second order perturbations from \texttt{RELIC}) and its representation in the particle-mesh scheme.}
\label{fig:maps100}
\end{figure}

To test the implementation in \texttt{gevolution} described in Section \ref{sec:gevolution}, we generate first- and second-order initial data with \texttt{RELIC} with the same box size and at the same redshift as used in the previous sections.
For this check, the fields are discretised on a $1024^3$ mesh and the second-order density is computed in Fourier space with \texttt{RELIC} only up to $25\%$ of the Nyquist wavenumber. As we will explain in Section \ref{sec:simulation}, we choose this setup in order to waste less computational resources for setting initial data at scales that will suffer severely from discretisation effects during the later evolution. We use Eqs.\ \eqref{eq:delta}, \eqref{eq:phi2} and \eqref{eq:velocitydivergence} to compute the density, the potential and the canonical momentum, respectively. Once \texttt{gevolution} has initialised the N-body particle ensemble (we use $2048^3$ particles for this test in order to guarantee a good sampling also in low-density regions) we obtain the density field from the particle-mesh projection, $\hat{\delta} = (\hat{\rho}/\bar{\rho}) - 1$, and compare it to the desired input field, $\delta = \delta^{(1)} + \delta^{(2)}$, from \texttt{RELIC}. In order to determine how well the method works at second order we compute $\hat{\delta} - \delta^{(1)}$ which we may call ``\texttt{gevolution} $\delta^{(2)}$'' and compare it to the input $\delta^{(2)}$. We also compute a residual field, $\hat{\delta} - \delta^{(1)} - \delta^{(2)}$. The power spectra of these fields are compared in Fig.\ \ref{fig:gevolution100} where we find excellent agreement between the input field and the result from the particle-mesh projection. The residual is indeed several orders of magnitude smaller than the second-order density which shows that our method is sufficiently accurate. A visual comparison of the different fields is presented in Fig.\ \ref{fig:maps100}, confirming that the agreement is not only statistical but also holds at the level of the realisation.

\subsection{Consistency of time evolution}

\label{sec:simulation}

\begin{figure}[tb]
\begin{center}
    \includegraphics[width=0.9\textwidth]{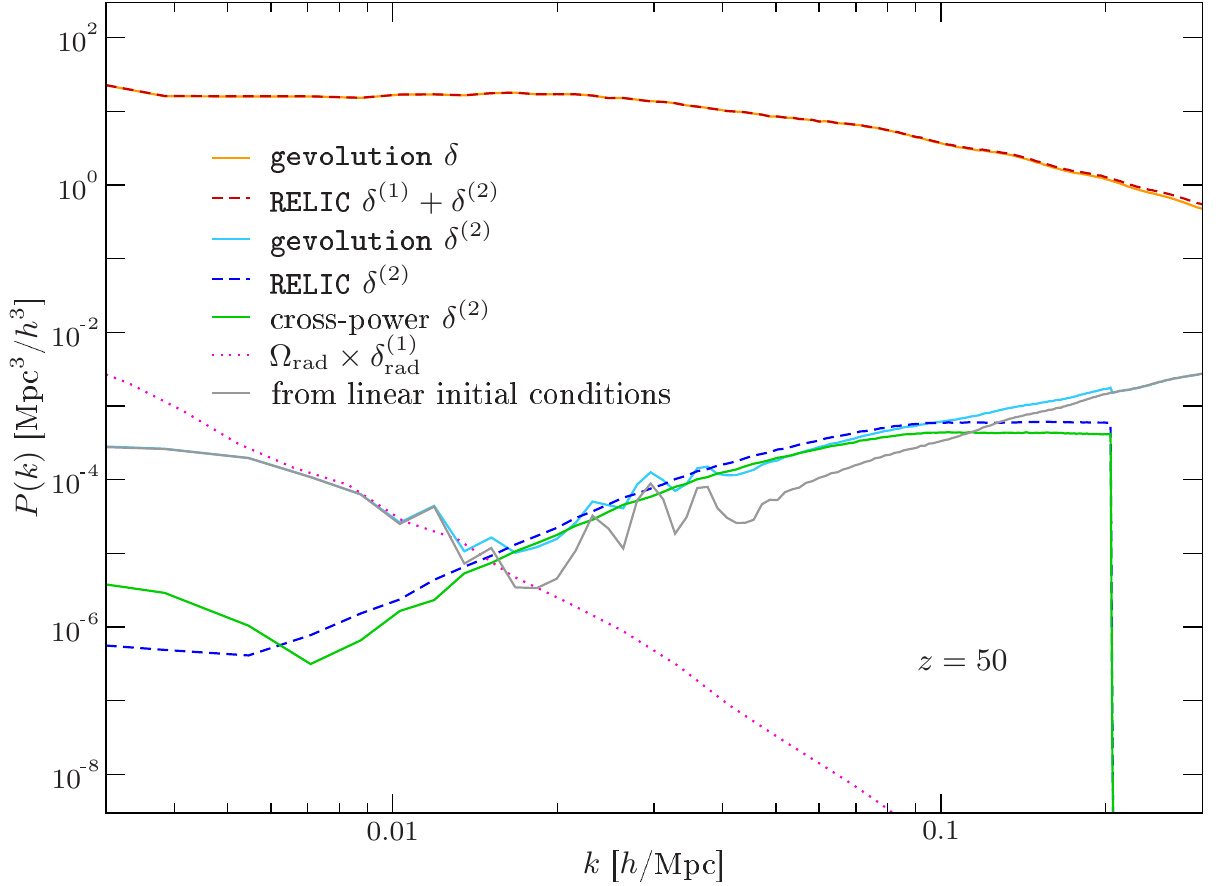}
\end{center}
    \caption{Power spectra of the density contrast at redshift $z=50$ in \texttt{RELIC} and in a simulation with \texttt{gevolution} that started from $z=100$. While the overall density contrast agrees very well as expected (orange and dashed red curves), the second-order density in the simulation (light blue curve) is affected by various numerical errors (see text). Nevertheless, we find good absolute agreement with the result from \texttt{RELIC} (dashed blue) on intermediate scales, and the cross-correlation (green) also reveals that $\delta^{(2)}$, while buried in errors, is still present at about the right amplitude on all other scales. We also show the second-order density in a simulation that started on linear initial conditions only (grey line), demonstrating further that the second-order contribution is significant. On large scales the effect of perturbations in the radiation field (pink dotted curve) is relevant even at redshift $z=50$.}
\label{fig:gevolution50}
\end{figure}

Within the regime where a second-order perturbative calculation is deemed accurate one would expect that the choice of initial redshift plays no role. Solving the mode evolution with a second-order Boltzmann code should be as good as doing it with an N-body code. We can check the consistency by setting up N-body initial data at some high redshift, evolving them with the N-body code to some lower redshift that is still high enough for a second-order calculation to remain valid, and comparing the solution with the one of the Boltzmann code for that same lower redshift value. However, there are (at least) two caveats to bear in mind.

First, a discrete system will always display a modified evolution with respect to the continuum limit as one approaches the discretisation scale, and this is true already at first order. Such a modified evolution will lead to a gradual departure from the continuum perturbation theory that, for some short scales, will become much larger than any second-order effects we are interested in. For the test we consider here, only scales that are about an order of magnitude or more above the Nyquist limit show a sufficient agreement with the continuum. This is the main reason why we choose not to compute the second-order density on the full Fourier grid, but only up to $25\%$ of the Nyquist wavenumber.

The second caveat concerns the physics modelled in the respective codes. For example, it is well known that the presence of perturbations in the radiation field (photons and neutrinos that are treated as massless here) has a small effect on the evolution of modes close to the horizon scale even during matter domination. While there are various ways to account for this effect in an N-body simulation (see Ref.\ \cite{Adamek:2017grt} for a fairly comprehensive discussion) the most common approach is to neglect those perturbations in the simulation and to ``fudge'' the first-order initial power spectrum such that the error becomes small at low redshift (this method is commonly known as ``backscaling''). At early times this can introduce spurious modifications on large scales that can easily dwarf the second-order relativistic effects we are interested in. This is clearly evident from Fig.\ \ref{fig:gevolution100} where we show the first-order perturbations in the radiation field (appropriately scaled by the respective density parameter) together with first- and second-order perturbations in the matter. To avoid this issue the most straightforward solution is to include the radiation field in the simulation, a feature that is conveniently available in \texttt{gevolution}. However, to employ this feature consistently we had to modify the code so that it uses the same random realisation that was provided by \texttt{RELIC} at initial time.

With these caveats in mind we run \texttt{gevolution} starting from redshift $z=100$ until the simulation reaches redshift $z=50$. We then compare the density perturbations in the simulation with the prediction of \texttt{RELIC} at the same redshift. In Fig.\ \ref{fig:gevolution50} we show the agreement of the power spectra of the total $\delta$ (orange solid and red dashed curves) as well as the agreement on the second-order contribution $\delta^{(2)}$ (light blue solid and blue dashed lines). To compute the latter quantity for the simulation we simply subtract $\delta^{(1)}$ (obtained with \texttt{RELIC}) from the simulation output. This means that the resulting field also contains some residuals from discretisation effects and other numerical errors. Therefore, in order to establish how well the fiducial $\delta^{(2)}$ is still captured in the simulation, we also show the cross-power between the simulation and the $\delta^{(2)}$ of \texttt{RELIC} (green solid curve). While we find results that are within $\sim 10\%$ agreement on intermediate scales, the small scales are clearly dominated by discretisation errors. There is also a considerable disagreement on very large scales. Looking at the power of radiation perturbations (pink dotted curve) it is tempting to conjecture that there is still some inconsistency in the treatment of the radiation effect. Note that radiation was included in the simulation using the method described in Ref.\ \cite{Adamek:2017grt}, and we checked that we get even poorer agreement if we neglect the effect of radiation perturbations. A deeper analysis is required to resolve this issue in the future.

We finally also compare our result at redshift $z=50$ with a simulation that used only linear initial conditions at redshift $z=100$. To this end we employ exactly the same simulation pipeline except that we set $\delta^{(2)} = 0$, $\phi^{(2)} = 0$, etc., initially. In Fig.\ \ref{fig:gevolution50} the grey curve indicates the perturbations recovered after subtracting $\delta^{(1)}$ from the final density contrast, as was done for the light blue curve. While the levels of contamination at large and small scales appear to be similar in both simulations, we confirm that the impact of second-order initial conditions can clearly be measured on intermediate scales.

\section{Conclusions}
\label{sec:concl}

N-body simulations of large-scale structure commonly use Newtonian gravity and Gaussian initial particle distributions. However, the increase of precision promised by the next generation of observational campaigns drives the community toward more accurate and complex numerical methods. Paving the way for ambitious non-Gaussian, relativistic N-body simulations, we developed \texttt{RELIC} to generate second-order initial Cauchy data. The core module of \texttt{RELIC} is an integrator that computes the convolution integral \eqref{eq:transfer22}. This integral is known to be computationally challenging for large grids. We have tailored an approximation to account for the coupling between large and small scales. We did this by ignoring the coupling between modes larger than a cut-off $k > k_\Lambda$. In this way, we correctly describe the so-called squeezed limit of the bispectrum. This limit is a primary probe of primordial non-Gaussianities as it is protected from many small-scales effects by virtue of the equivalence principle\ \cite{Creminelli:2013mca}. Our approximation also correctly describes the full bispectrum at large scales $k < k_\Lambda$. 

We detailed how our approximation reduces the computational time. We checked that the power spectra and bispectra of the realisations generated by \texttt{RELIC} are indeed the ones provided by \texttt{CLASS} and \texttt{SONG}. Passing the Cauchy data to \texttt{gevolution}, we also checked that the particle-mesh projection nicely reproduces the \texttt{RELIC} density by visual inspection and measurements of the power spectrum. Finally, we checked in that the time evolution of the perturbations is consistent with perturbation theory from redshift $z=100$ to $z=50$, except for some systematic residual on very large scales that remains to be fully understood.

Though our focus was on relativistic simulations, \texttt{RELIC} is a general-purpose tool to generate initial conditions with an arbitrary coupling between long and short scales. It is straightforward to include a PNG signal, and \texttt{SONG} already provides some means to do that. In fact, \texttt{RELIC} can work with arbitrary second-order kernels. Apart from our implementation in \texttt{gevolution}, the Cauchy data prepared by the generator can also be used by other simulation pipelines that accept fields as an input, e.g.\ the \texttt{Einstein toolkit} \cite{Zilhao:2013hia}.

In order to go beyond local PNG and include the most general (non-separable) non-Gaussian template, the modal approach discussed in Section \ref{sec:Introduction} could be applied. As one approaches the horizon scale, the inclusion of the value of the density and velocity fields at the fundamental mode of the simulation becomes highly relevant. We leave these considerations for future works.

Another virtue of considering non-linear initial conditions is that it allows to start the N-body simulation at later redshift, thus eliminating transient effects, see Ref.\ \cite{Michaux:2020yis} for an explicit example at 3LPT. In follow-up work, we also want to use another in-built feature of the code we are working with, namely the ray-tracing algorithm of \texttt{gevolution} (see for instance Ref.\ \cite{Lepori:2020ifz}), that combined with our current pipeline, would allow to make a prediction for the observable galaxy power spectrum and bispectrum accounting for all relativistic and non-linear effects. Being able to disentangle relativistic effects and the primordial signal would be of major interest for future observational surveys.

\acknowledgments
We thank Matteo Biagetti for helpful discussions. Our numerical work was facilitated by the IN2P3 Computing Centre
(\url{https://cc.in2p3.fr}) and the ScienceCluster of Service and Support for Science IT (S${}^\text{3}$IT) at the University of Zurich. We acknowledge the use of the Python library \texttt{Pylians3} (\url{https://pylians3.readthedocs.io}) for the computation of power spectra and bispectra.  JA acknowledges funding by the Swiss National Science Foundation. JC is supported by ANID scholarship No.\ 21210008. TM thanks the Institute for Computational Science for hospitality and the Paris Centre for Cosmological Physics for financial support. JN is supported by FONDECYT grant 1211545, ``Measuring the Field Spectrum of the Early Universe''. CS acknowledges funding from the European Research Council (ERC) under the European Union's Horizon 2020 research and innovation programme (grant agreement No.\ 834148).

\bibliographystyle{JHEP.bst}
\bibliography{main.bib}

\providecommand{\href}[2]{#2}\begingroup\raggedright\begin{thebibliography}{10}

\bibitem{Maldacena:2002vr}
J.~M. Maldacena, \emph{{Non-Gaussian features of primordial fluctuations in
  single field inflationary models}},
  \href{http://dx.doi.org/10.1088/1126-6708/2003/05/013}{\emph{JHEP} {\bf 05}
  (2003) 013}, [\href{https://arxiv.org/abs/astro-ph/0210603}{{\tt
  astro-ph/0210603}}].

\bibitem{Creminelli:2004yq}
P.~Creminelli and M.~Zaldarriaga, \emph{{Single field consistency relation for
  the 3-point function}},
  \href{http://dx.doi.org/10.1088/1475-7516/2004/10/006}{\emph{JCAP} {\bf 10}
  (2004) 006}, [\href{https://arxiv.org/abs/astro-ph/0407059}{{\tt
  astro-ph/0407059}}].

\bibitem{Creminelli:2011rh}
P.~Creminelli, G.~D'Amico, M.~Musso and J.~Nore\~na, \emph{{The (not so)
  squeezed limit of the primordial 3-point function}},
  \href{http://dx.doi.org/10.1088/1475-7516/2011/11/038}{\emph{JCAP} {\bf 11}
  (2011) 038}, [\href{https://arxiv.org/abs/1106.1462}{{\tt 1106.1462}}].

\bibitem{Creminelli:2012ed}
P.~Creminelli, J.~Nore\~na and M.~Simonovi\'c, \emph{{Conformal consistency
  relations for single-field inflation}},
  \href{http://dx.doi.org/10.1088/1475-7516/2012/07/052}{\emph{JCAP} {\bf 07}
  (2012) 052}, [\href{https://arxiv.org/abs/1203.4595}{{\tt 1203.4595}}].

\bibitem{Pajer:2013ana}
E.~Pajer, F.~Schmidt and M.~Zaldarriaga, \emph{{The Observed Squeezed Limit of
  Cosmological Three-Point Functions}},
  \href{http://dx.doi.org/10.1103/PhysRevD.88.083502}{\emph{Phys. Rev. D} {\bf
  88} (2013) 083502}, [\href{https://arxiv.org/abs/1305.0824}{{\tt
  1305.0824}}].

\bibitem{Gangui:1993tt}
A.~Gangui, F.~Lucchin, S.~Matarrese and S.~Mollerach, \emph{{The Three point
  correlation function of the cosmic microwave background in inflationary
  models}}, \href{http://dx.doi.org/10.1086/174421}{\emph{Astrophys. J.} {\bf
  430} (1994) 447--457}, [\href{https://arxiv.org/abs/astro-ph/9312033}{{\tt
  astro-ph/9312033}}].

\bibitem{Komatsu:2010hc}
E.~Komatsu, \emph{{Hunting for Primordial Non-Gaussianity in the Cosmic
  Microwave Background}},
  \href{http://dx.doi.org/10.1088/0264-9381/27/12/124010}{\emph{Class. Quant.
  Grav.} {\bf 27} (2010) 124010}, [\href{https://arxiv.org/abs/1003.6097}{{\tt
  1003.6097}}].

\bibitem{Akrami:2019izv}
{\scshape Planck} collaboration, Y.~Akrami et~al., \emph{{Planck 2018 results.
  IX. Constraints on primordial non-Gaussianity}},
  \href{http://dx.doi.org/10.1051/0004-6361/201935891}{\emph{Astron.
  Astrophys.} {\bf 641} (2020) A9},
  [\href{https://arxiv.org/abs/1905.05697}{{\tt 1905.05697}}].

\bibitem{Mueller:2021tqa}
E.-M. Mueller et~al., \emph{{The clustering of galaxies in the completed
  SDSS-IV extended Baryon Oscillation Spectroscopic Survey: Primordial
  non-Gaussianity in Fourier Space}},
  \href{https://arxiv.org/abs/2106.13725}{{\tt 2106.13725}}.

\bibitem{Amendola:2016saw}
L.~Amendola et~al., \emph{{Cosmology and fundamental physics with the Euclid
  satellite}}, \href{http://dx.doi.org/10.1007/s41114-017-0010-3}{\emph{Living
  Rev. Rel.} {\bf 21} (2018) 2}, [\href{https://arxiv.org/abs/1606.00180}{{\tt
  1606.00180}}].

\bibitem{Zhan:2017uwu}
H.~Zhan and J.~A. Tyson, \emph{{Cosmology with the Large Synoptic Survey
  Telescope: an Overview}},
  \href{http://dx.doi.org/10.1088/1361-6633/aab1bd}{\emph{Rept. Prog. Phys.}
  {\bf 81} (2018) 066901}, [\href{https://arxiv.org/abs/1707.06948}{{\tt
  1707.06948}}].

\bibitem{SKA:2018ckk}
{\scshape SKA} collaboration, D.~J. Bacon et~al., \emph{{Cosmology with Phase 1
  of the Square Kilometre Array: Red Book 2018: Technical specifications and
  performance forecasts}},
  \href{http://dx.doi.org/10.1017/pasa.2019.51}{\emph{Publ. Astron. Soc.
  Austral.} {\bf 37} (2020) e007},
  [\href{https://arxiv.org/abs/1811.02743}{{\tt 1811.02743}}].

\bibitem{Dore:2014cca}
O.~Dor\'e et~al., \emph{{Cosmology with the SPHEREX All-Sky Spectral Survey}},
  \href{https://arxiv.org/abs/1412.4872}{{\tt 1412.4872}}.

\bibitem{Schneider:2015yka}
A.~Schneider, R.~Teyssier, D.~Potter, J.~Stadel, J.~Onions, D.~S. Reed et~al.,
  \emph{{Matter power spectrum and the challenge of percent accuracy}},
  \href{http://dx.doi.org/10.1088/1475-7516/2016/04/047}{\emph{JCAP} {\bf 04}
  (2016) 047}, [\href{https://arxiv.org/abs/1503.05920}{{\tt 1503.05920}}].

\bibitem{Biagetti:2019bnp}
M.~Biagetti, \emph{{The Hunt for Primordial Interactions in the Large Scale
  Structures of the Universe}},
  \href{http://dx.doi.org/10.3390/galaxies7030071}{\emph{Galaxies} {\bf 7}
  (2019) 71}, [\href{https://arxiv.org/abs/1906.12244}{{\tt 1906.12244}}].

\bibitem{Desjacques:2016bnm}
V.~Desjacques, D.~Jeong and F.~Schmidt, \emph{{Large-Scale Galaxy Bias}},
  \href{http://dx.doi.org/10.1016/j.physrep.2017.12.002}{\emph{Phys. Rept.}
  {\bf 733} (2018) 1--193}, [\href{https://arxiv.org/abs/1611.09787}{{\tt
  1611.09787}}].

\bibitem{Karagiannis:2018jdt}
D.~Karagiannis, A.~Lazanu, M.~Liguori, A.~Raccanelli, N.~Bartolo and L.~Verde,
  \emph{{Constraining primordial non-Gaussianity with bispectrum and power
  spectrum from upcoming optical and radio surveys}},
  \href{http://dx.doi.org/10.1093/mnras/sty1029}{\emph{Mon. Not. Roy. Astron.
  Soc.} {\bf 478} (2018) 1341--1376},
  [\href{https://arxiv.org/abs/1801.09280}{{\tt 1801.09280}}].

\bibitem{Karagiannis:2019jjx}
D.~Karagiannis, A.~Slosar and M.~Liguori, \emph{{Forecasts on Primordial
  non-Gaussianity from 21 cm Intensity Mapping experiments}},
  \href{http://dx.doi.org/10.1088/1475-7516/2020/11/052}{\emph{JCAP} {\bf 11}
  (2020) 052}, [\href{https://arxiv.org/abs/1911.03964}{{\tt 1911.03964}}].

\bibitem{Bernardeau:2001qr}
F.~Bernardeau, S.~Colombi, E.~Gazta\~naga and R.~Scoccimarro, \emph{{Large
  scale structure of the universe and cosmological perturbation theory}},
  \href{http://dx.doi.org/10.1016/S0370-1573(02)00135-7}{\emph{Phys. Rept.}
  {\bf 367} (2002) 1--248}, [\href{https://arxiv.org/abs/astro-ph/0112551}{{\tt
  astro-ph/0112551}}].

\bibitem{Matsubara:1995kq}
T.~Matsubara, \emph{{On second order perturbation theories of gravitational
  instability in Friedmann-Lemaitre models}},
  \href{http://dx.doi.org/10.1143/PTP.94.1151}{\emph{Prog. Theor. Phys.} {\bf
  94} (1995) 1151--1156}, [\href{https://arxiv.org/abs/astro-ph/9510137}{{\tt
  astro-ph/9510137}}].

\bibitem{Matarrese:1997ay}
S.~Matarrese, S.~Mollerach and M.~Bruni, \emph{{Second order perturbations of
  the Einstein-de Sitter universe}},
  \href{http://dx.doi.org/10.1103/PhysRevD.58.043504}{\emph{Phys. Rev. D} {\bf
  58} (1998) 043504}, [\href{https://arxiv.org/abs/astro-ph/9707278}{{\tt
  astro-ph/9707278}}].

\bibitem{Boubekeur:2008kn}
L.~Boubekeur, P.~Creminelli, J.~Nore\~na and F.~Vernizzi, \emph{{Action
  approach to cosmological perturbations: the 2nd order metric in matter
  dominance}},
  \href{http://dx.doi.org/10.1088/1475-7516/2008/08/028}{\emph{JCAP} {\bf 08}
  (2008) 028}, [\href{https://arxiv.org/abs/0806.1016}{{\tt 0806.1016}}].

\bibitem{Bartolo:2010rw}
N.~Bartolo, S.~Matarrese, O.~Pantano and A.~Riotto, \emph{{Second-order matter
  perturbations in a LambdaCDM cosmology and non-Gaussianity}},
  \href{http://dx.doi.org/10.1088/0264-9381/27/12/124009}{\emph{Class. Quant.
  Grav.} {\bf 27} (2010) 124009}, [\href{https://arxiv.org/abs/1002.3759}{{\tt
  1002.3759}}].

\bibitem{Bruni:2013qta}
M.~Bruni, J.~C. Hidalgo, N.~Meures and D.~Wands, \emph{{Non-Gaussian Initial
  Conditions in \ensuremath{\Lambda}CDM: Newtonian, Relativistic, and
  Primordial Contributions}},
  \href{http://dx.doi.org/10.1088/0004-637X/785/1/2}{\emph{Astrophys. J.} {\bf
  785} (2014) 2}, [\href{https://arxiv.org/abs/1307.1478}{{\tt 1307.1478}}].

\bibitem{Villa:2014foa}
E.~Villa, L.~Verde and S.~Matarrese, \emph{{General relativistic corrections
  and non-Gaussianity in large scale structure}},
  \href{http://dx.doi.org/10.1088/0264-9381/31/23/234005}{\emph{Class. Quant.
  Grav.} {\bf 31} (2014) 234005}, [\href{https://arxiv.org/abs/1409.4738}{{\tt
  1409.4738}}].

\bibitem{Fitzpatrick:2009ci}
A.~L. Fitzpatrick, L.~Senatore and M.~Zaldarriaga, \emph{{Contributions to the
  dark matter 3-Point function from the radiation era}},
  \href{http://dx.doi.org/10.1088/1475-7516/2010/05/004}{\emph{JCAP} {\bf 05}
  (2010) 004}, [\href{https://arxiv.org/abs/0902.2814}{{\tt 0902.2814}}].

\bibitem{Tram:2016cpy}
T.~Tram, C.~Fidler, R.~Crittenden, K.~Koyama, G.~W. Pettinari and D.~Wands,
  \emph{{The Intrinsic Matter Bispectrum in $\Lambda$CDM}},
  \href{http://dx.doi.org/10.1088/1475-7516/2016/05/058}{\emph{JCAP} {\bf 05}
  (2016) 058}, [\href{https://arxiv.org/abs/1602.05933}{{\tt 1602.05933}}].

\bibitem{Villa:2015ppa}
E.~Villa and C.~Rampf, \emph{{Relativistic perturbations in $\Lambda$CDM:
  Eulerian \& Lagrangian approaches}},
  \href{http://dx.doi.org/10.1088/1475-7516/2016/01/030}{\emph{JCAP} {\bf 01}
  (2016) 030}, [\href{https://arxiv.org/abs/1505.04782}{{\tt 1505.04782}}].

\bibitem{DiDio:2016gpd}
E.~Di~Dio, H.~Perrier, R.~Durrer, G.~Marozzi, A.~Moradinezhad~Dizgah,
  J.~Nore\~na et~al., \emph{{Non-Gaussianities due to Relativistic Corrections
  to the Observed Galaxy Bispectrum}},
  \href{http://dx.doi.org/10.1088/1475-7516/2017/03/006}{\emph{JCAP} {\bf 03}
  (2017) 006}, [\href{https://arxiv.org/abs/1611.03720}{{\tt 1611.03720}}].

\bibitem{Castiblanco:2018qsd}
L.~Castiblanco, R.~Gannouji, J.~Nore\~na and C.~Stahl, \emph{{Relativistic
  cosmological large scale structures at one-loop}},
  \href{http://dx.doi.org/10.1088/1475-7516/2019/07/030}{\emph{JCAP} {\bf 07}
  (2019) 030}, [\href{https://arxiv.org/abs/1811.05452}{{\tt 1811.05452}}].

\bibitem{Kehagias:2015tda}
A.~Kehagias, A.~Moradinezhad~Dizgah, J.~Nore\~na, H.~Perrier and A.~Riotto,
  \emph{{A Consistency Relation for the Observed Galaxy Bispectrum and the
  Local non-Gaussianity from Relativistic Corrections}},
  \href{http://dx.doi.org/10.1088/1475-7516/2015/08/018}{\emph{JCAP} {\bf 08}
  (2015) 018}, [\href{https://arxiv.org/abs/1503.04467}{{\tt 1503.04467}}].

\bibitem{Bartolo:2015qva}
N.~Bartolo, D.~Bertacca, M.~Bruni, K.~Koyama, R.~Maartens, S.~Matarrese et~al.,
  \emph{{A relativistic signature in large-scale structure}},
  \href{http://dx.doi.org/10.1016/j.dark.2016.04.002}{\emph{Phys. Dark Univ.}
  {\bf 13} (2016) 30--34}, [\href{https://arxiv.org/abs/1506.00915}{{\tt
  1506.00915}}].

\bibitem{Creminelli:2013mca}
P.~Creminelli, J.~Nore\~na, M.~Simonovi\'c and F.~Vernizzi, \emph{{Single-Field
  Consistency Relations of Large Scale Structure}},
  \href{http://dx.doi.org/10.1088/1475-7516/2013/12/025}{\emph{JCAP} {\bf 12}
  (2013) 025}, [\href{https://arxiv.org/abs/1309.3557}{{\tt 1309.3557}}].

\bibitem{dePutter:2015vga}
R.~de~Putter, O.~Dor\'e and D.~Green, \emph{{Is There Scale-Dependent Bias in
  Single-Field Inflation?}},
  \href{http://dx.doi.org/10.1088/1475-7516/2015/10/024}{\emph{JCAP} {\bf 10}
  (2015) 024}, [\href{https://arxiv.org/abs/1504.05935}{{\tt 1504.05935}}].

\bibitem{Pitrou:2008ut}
C.~Pitrou, \emph{{The radiative transfer for polarized radiation at second
  order in cosmological perturbations}},
  \href{http://dx.doi.org/10.1007/s10714-009-0782-1}{\emph{Gen. Rel. Grav.}
  {\bf 41} (2009) 2587--2595}, [\href{https://arxiv.org/abs/0809.3245}{{\tt
  0809.3245}}].

\bibitem{Pettinari:2014vja}
G.~W. Pettinari, \emph{{The intrinsic bispectrum of the Cosmic Microwave
  Background}}.
\newblock PhD thesis, Portsmouth U., ICG, 9, 2013.
\newblock \href{https://arxiv.org/abs/1405.2280}{{\tt 1405.2280}}.
\newblock 10.1007/978-3-319-21882-3.

\bibitem{Adamek:2015eda}
J.~Adamek, D.~Daverio, R.~Durrer and M.~Kunz, \emph{{General relativity and
  cosmic structure formation}},
  \href{http://dx.doi.org/10.1038/nphys3673}{\emph{Nature Phys.} {\bf 12}
  (2016) 346--349}, [\href{https://arxiv.org/abs/1509.01699}{{\tt
  1509.01699}}].

\bibitem{Adamek:2016zes}
J.~Adamek, D.~Daverio, R.~Durrer and M.~Kunz, \emph{{gevolution: a cosmological
  N-body code based on General Relativity}},
  \href{http://dx.doi.org/10.1088/1475-7516/2016/07/053}{\emph{JCAP} {\bf 07}
  (2016) 053}, [\href{https://arxiv.org/abs/1604.06065}{{\tt 1604.06065}}].

\bibitem{Wagner:2010me}
C.~Wagner, L.~Verde and L.~Boubekeur, \emph{{N-body simulations with generic
  non-Gaussian initial conditions I: Power Spectrum and halo mass function}},
  \href{http://dx.doi.org/10.1088/1475-7516/2010/10/022}{\emph{JCAP} {\bf 10}
  (2010) 022}, [\href{https://arxiv.org/abs/1006.5793}{{\tt 1006.5793}}].

\bibitem{Wagner:2011wx}
C.~Wagner and L.~Verde, \emph{{N-body simulations with generic non-Gaussian
  initial conditions II: Halo bias}},
  \href{http://dx.doi.org/10.1088/1475-7516/2012/03/002}{\emph{JCAP} {\bf 03}
  (2012) 002}, [\href{https://arxiv.org/abs/1102.3229}{{\tt 1102.3229}}].

\bibitem{Scoccimarro:2011pz}
R.~Scoccimarro, L.~Hui, M.~Manera and K.~C. Chan, \emph{{Large-scale Bias and
  Efficient Generation of Initial Conditions for Non-Local Primordial
  Non-Gaussianity}},
  \href{http://dx.doi.org/10.1103/PhysRevD.85.083002}{\emph{Phys. Rev. D} {\bf
  85} (2012) 083002}, [\href{https://arxiv.org/abs/1108.5512}{{\tt
  1108.5512}}].

\bibitem{Adhikari:2014xua}
S.~Adhikari, S.~Shandera and N.~Dalal, \emph{{Higher moments of primordial
  non-Gaussianity and N-body simulations}},
  \href{http://dx.doi.org/10.1088/1475-7516/2014/06/052}{\emph{JCAP} {\bf 06}
  (2014) 052}, [\href{https://arxiv.org/abs/1402.2336}{{\tt 1402.2336}}].

\bibitem{Smith:2010gx}
K.~M. Smith and M.~LoVerde, \emph{{Local stochastic non-Gaussianity and N-body
  simulations}},
  \href{http://dx.doi.org/10.1088/1475-7516/2011/11/009}{\emph{JCAP} {\bf 11}
  (2011) 009}, [\href{https://arxiv.org/abs/1010.0055}{{\tt 1010.0055}}].

\bibitem{Regan:2011zq}
D.~M. Regan, M.~M. Schmittfull, E.~P.~S. Shellard and J.~R. Fergusson,
  \emph{{Universal Non-Gaussian Initial Conditions for N-body Simulations}},
  \href{http://dx.doi.org/10.1103/PhysRevD.86.123524}{\emph{Phys. Rev. D} {\bf
  86} (2012) 123524}, [\href{https://arxiv.org/abs/1108.3813}{{\tt
  1108.3813}}].

\bibitem{Fergusson:2010dm}
J.~R. Fergusson, M.~Liguori and E.~P.~S. Shellard, \emph{{The CMB Bispectrum}},
  \href{http://dx.doi.org/10.1088/1475-7516/2012/12/032}{\emph{JCAP} {\bf 12}
  (2012) 032}, [\href{https://arxiv.org/abs/1006.1642}{{\tt 1006.1642}}].

\bibitem{Fergusson:2010ia}
J.~R. Fergusson, D.~M. Regan and E.~P.~S. Shellard, \emph{{Rapid Separable
  Analysis of Higher Order Correlators in Large Scale Structure}},
  \href{http://dx.doi.org/10.1103/PhysRevD.86.063511}{\emph{Phys. Rev. D} {\bf
  86} (2012) 063511}, [\href{https://arxiv.org/abs/1008.1730}{{\tt
  1008.1730}}].

\bibitem{Hung:2019ygc}
J.~Hung, J.~R. Fergusson and E.~P.~S. Shellard, \emph{{Advancing the matter
  bispectrum estimation of large-scale structure: a comparison of dark matter
  codes}},  \href{https://arxiv.org/abs/1902.01830}{{\tt 1902.01830}}.

\bibitem{Enriquez:2021arn}
M.~Enr\'\i{}quez, J.~C. Hidalgo and O.~Valenzuela, \emph{{Cosmological
  simulations with relativistic and primordial non-Gaussianity contributions as
  initial conditions}},  \href{https://arxiv.org/abs/2109.13364}{{\tt
  2109.13364}}.

\bibitem{Baumann:2009ds}
D.~Baumann, \emph{{Inflation}},  in \emph{{Theoretical Advanced Study Institute
  in Elementary Particle Physics}: {Physics of the Large and the Small}},
  pp.~523--686, 2011.
\newblock \href{https://arxiv.org/abs/0907.5424}{{\tt 0907.5424}}.
\newblock \href{http://dx.doi.org/10.1142/9789814327183_0010}{DOI}.

\bibitem{Pitrou:2010sn}
C.~Pitrou, J.-P. Uzan and F.~Bernardeau, \emph{{The cosmic microwave background
  bispectrum from the non-linear evolution of the cosmological perturbations}},
  \href{http://dx.doi.org/10.1088/1475-7516/2010/07/003}{\emph{JCAP} {\bf 07}
  (2010) 003}, [\href{https://arxiv.org/abs/1003.0481}{{\tt 1003.0481}}].

\bibitem{Blas:2011rf}
D.~Blas, J.~Lesgourgues and T.~Tram, \emph{{The Cosmic Linear Anisotropy
  Solving System (CLASS) II: Approximation schemes}},
  \href{http://dx.doi.org/10.1088/1475-7516/2011/07/034}{\emph{JCAP} {\bf 07}
  (2011) 034}, [\href{https://arxiv.org/abs/1104.2933}{{\tt 1104.2933}}].

\bibitem{Lu:2008ju}
T.~H.-C. Lu, K.~Ananda, C.~Clarkson and R.~Maartens, \emph{{The cosmological
  background of vector modes}},
  \href{http://dx.doi.org/10.1088/1475-7516/2009/02/023}{\emph{JCAP} {\bf 02}
  (2009) 023}, [\href{https://arxiv.org/abs/0812.1349}{{\tt 0812.1349}}].

\bibitem{Planck:2013pxb}
{\scshape Planck} collaboration, P.~A.~R. Ade et~al., \emph{{Planck 2013
  results. XVI. Cosmological parameters}},
  \href{http://dx.doi.org/10.1051/0004-6361/201321591}{\emph{Astron.
  Astrophys.} {\bf 571} (2014) A16},
  [\href{https://arxiv.org/abs/1303.5076}{{\tt 1303.5076}}].

\bibitem{Lewandowski:2017kes}
M.~Lewandowski and L.~Senatore, \emph{{IR-safe and UV-safe integrands in the
  EFTofLSS with exact time dependence}},
  \href{http://dx.doi.org/10.1088/1475-7516/2017/08/037}{\emph{JCAP} {\bf 08}
  (2017) 037}, [\href{https://arxiv.org/abs/1701.07012}{{\tt 1701.07012}}].

\bibitem{Adamek:2017grt}
J.~Adamek, J.~Brandbyge, C.~Fidler, S.~Hannestad, C.~Rampf and T.~Tram,
  \emph{{The effect of early radiation in N-body simulations of cosmic
  structure formation}},
  \href{http://dx.doi.org/10.1093/mnras/stx1157}{\emph{Mon. Not. Roy. Astron.
  Soc.} {\bf 470} (2017) 303--313},
  [\href{https://arxiv.org/abs/1703.08585}{{\tt 1703.08585}}].

\bibitem{Zilhao:2013hia}
M.~Zilh\~ao and F.~L\"offler, \emph{{An Introduction to the Einstein Toolkit}},
  \href{http://dx.doi.org/10.1142/S0217751X13400149}{\emph{Int. J. Mod. Phys.
  A} {\bf 28} (2013) 1340014}, [\href{https://arxiv.org/abs/1305.5299}{{\tt
  1305.5299}}].

\bibitem{Michaux:2020yis}
M.~Michaux, O.~Hahn, C.~Rampf and R.~E. Angulo, \emph{{Accurate initial
  conditions for cosmological N-body simulations: Minimizing truncation and
  discreteness errors}},
  \href{http://dx.doi.org/10.1093/mnras/staa3149}{\emph{Mon. Not. Roy. Astron.
  Soc.} {\bf 500} (2020) 663--683},
  [\href{https://arxiv.org/abs/2008.09588}{{\tt 2008.09588}}].

\bibitem{Lepori:2020ifz}
F.~Lepori, J.~Adamek, R.~Durrer, C.~Clarkson and L.~Coates, \emph{{Weak-lensing
  observables in relativistic N-body simulations}},
  \href{http://dx.doi.org/10.1093/mnras/staa2024}{\emph{Mon. Not. Roy. Astron.
  Soc.} {\bf 497} (2020) 2078--2095},
  [\href{https://arxiv.org/abs/2002.04024}{{\tt 2002.04024}}].

\end{thebibliography}\endgroup

\end{document}